\documentclass[10pt,journal,compsoc]{IEEEtran}
\IEEEoverridecommandlockouts
\usepackage{cite}
\usepackage{amsmath,amssymb,amsfonts}
\usepackage{algorithmic}
\usepackage{graphicx}
\usepackage{textcomp}
\usepackage{xcolor}

\usepackage{subfig}
\newtheorem{theorem}{Theorem}  [section]

\def\BibTeX{{\rm B\kern-.05em{\sc i\kern-.025em b}\kern-.08em
    T\kern-.1667em\lower.7ex\hbox{E}\kern-.125emX}}
\begin{document}

\title{Egoistic Incentives Based on Zero-Determinant Alliances for Large-Scale Systems\\
}
\author{Shengling Wang,~\IEEEmembership{Member,~IEEE,} Peizi Ma,~Qin Hu,~Xiuzhen Cheng,~\IEEEmembership{Fellow,~IEEE,} and Weifeng Lv 
\IEEEcompsocitemizethanks{
\IEEEcompsocthanksitem Shengling Wang and Peizi Ma are with the School of Artificial Intelligence, Beijing Normal University, Beijing, China.\protect\\
E-mail: wangshengling@bnu.edu.cn, mapeizi@mail.bnu.edu.cn
\IEEEcompsocthanksitem Qin Hu is with the 
Department of Computer and Information Science, Indiana University - Purdue University Indianapolis, IN, USA.\protect\\
E-mail: qinhu@iu.edu
\IEEEcompsocthanksitem Xiuzhen Cheng is with the Department of Computer Science, The George Washington University, Washington DC, USA.\protect\\
E-mail: cheng@gwu.edu 
\IEEEcompsocthanksitem Weifeng Lv (corresponding author) is with the School of Computer Science and Engineering, Beihang University, Beijing, China.\protect\\
E-mail: lwf@buaa.edu.cn
}
}


\IEEEtitleabstractindextext{
\begin{abstract}
Social dilemmas exist in various fields and give rise to the so-called {\it free-riding} problem, leading to collective fiascos. The difficulty of tracking individual behaviors makes egoistic incentives in large-scale systems a challenging task. However, the state-of-the-art mechanisms are either individual-based or state-dependent, resulting in low efficiency in large-scale networks. In this paper, we propose an egoistic incentive mechanism from  a {\it connected} (network) perspective rather than an {\it isolated} (individual) perspective by taking advantage of the social nature of people. We make use of a zero-determinant (ZD) strategy for rewarding cooperation and sanctioning defection. After proving cooperation is  the dominant strategy for ZD players, we optimize their deployment to facilitate cooperation over the whole system.  To further speed up cooperation,   we  derive a ZD alliance strategy for  sequential multiple-player repeated games to  empower ZD players with higher controllable leverage, which undoubtedly enriches the theoretical system of ZD strategies and broadens their application domain. Our approach is stateless and stable, which contribute to its scalability. Extensive simulations based on a real world trace data as well as synthetic data demonstrate the effectiveness of our proposed egoistic incentive approach under different networking scenarios.
\end{abstract}

\begin{IEEEkeywords}
Social dilemma, egoistic incentives, zero-determinant alliances.
\end{IEEEkeywords}}

\maketitle

\section{Introduction}
\label{sec:intro}

Many human decisions occur in situations where the results of one's own decisions are interdependent with those of others. Such interdependence situations tend to breed social dilemmas, which has two traits: 1) each individual who makes social defective choices gains higher profits no matter what  other individuals do; 2) comparing with the situation where everyone cooperates, if everyone chooses to defect,  then all individuals get lower returns \cite{Dawes1980Social}.  Therefore, a social dilemma is essentially an inherent conflict between defection and cooperation in a system  where the former is the dominant strategy for each individual but the latter can maximize the overall social welfare.

Social dilemmas exist in various scenarios such as data delivery in mobile opportunistic networks \cite{wangicdcs2017},  file sharing in peer-to-peer networks \cite{hu2009budget}, and autonomous vehicle programming \cite{science}. A typical issue resulted from a social dilemma  is {\it free-riding}, an individually rational but socially defecting choice which can lead to a collective fiasco. Moreover, such a {\it tragedy of the commons} would be aggravated as the system expands because the behavior of an individual becomes more difficult to track so that the influence on others diminishes \cite{pnus}.
Hence, there is a pressing need  to induce cooperation,  especially in a large-scale system.

Cooperation in a social dilemma  is often explained in terms of {\it  egoistic incentives} \cite{zhao}. The state-of-the-art egoistic incentive mechanisms realize their aims by transforming a social dilemma game into one not involving a dilemma,  which can be categorized into two types: reputation-based \cite{Lu2010Pi,uddin2010relics,hu2009budget,chen2015multicent,Ning15} and pricing-based \cite{ning13,chen,Koutsopoulos,Dejunmobicom}. Reputation-based schemes employ reputation or quasi-reputation to  evaluate a node's contribution  to others, based on which reacts to its service requests, while pricing-based mechanisms treat service provision as a transaction, taking advantage of monetary incentives to stimulate cooperation.

Existing incentive mechanisms have two common traits: they all are \emph{individual-based} and involve \emph{state (reputation or transaction status)  maintenance and management}. The first trait comes from the underlying concept in which the success of cooperation inducement to each individual inevitably results in that of the whole system. Such a case-by-case based approach is obviously inefficient as the size of the system enlarges. The second trait exists due to the state-dependent nature of the state-of-the-art methods. Regretfully, the cost of  maintaining and managing states would soar in a large-scale system.

In this paper, we realize large-scale egoistic incentives from the perspective of {\it  the  network} rather than  {\it an individual}.  Our idea  stems from the following observation:  each person in this world  is more or less  socially connected to others,  forming a so-called {\it  social community}. Within such an environment,  cooperation and competition among all members jointly determine the utility of an individual. Since each individual is utility-driven, cooperation inducement requires us to analyze the game among all players in the social community, which involves the structure of their social ties. Hence, it is reasonable to consider incentives from the {\it connected} (network) perspective rather than the {\it isolated} (individual) perspective to stimulate cooperation. Moreover,  directly incentivizing the whole system has the potential to bring high efficiency to a large-scale system.

Our egoistic incentive approach has two desired properties: {\it statelessness} and {\it stability}.  These  merits are  obtained by taking advantage of  the zero-determinant (ZD) strategy \cite{press2012iterated} whose adopter (the ZD  player) can control its  opponent's payoff  in a unilateral way. Thus, through rewarding cooperation and negatively sanctioning defection,  a ZD player can stimulate cooperation of its co-players. After proving cooperation is  the dominant strategy of the ZD players, we optimize their deployment, facilitating cooperation over the whole system. Our method is stateless because it does not need to manage or maintain any state; it is stable since the deployment of the ZD players only depends on the social ties among the players, which are usually steady. The statelessness and stability contribute to the sound scalability of our egoistic incentive.

Another contribution of this paper is the derivation of a ZD alliance strategy for  sequential multiple-player repeated games, where a ZD alliance refers to a group of players taking the same ZD strategy. Such a derivation brings two benefits. First, the alliance strategy widens the  range of a co-player's utility that ZD players can set, implying that ZD players can achieve higher controllable leverage  through allying. The increased controlling power of a ZD alliance breeds  an environment where cooperation thrives. Second, our derivation enriches the theoretical system of ZD strategies, broadening their application domain.

We conduct extensive simulations based on  real-world trace data as well as synthetic data, which represent different types of typical network topologies including  {\it star}, {\it ring}, {\it tree}, and {\it mesh} structures.  These simulation results demonstrate the effectiveness of our  egoistic incentive approach.

The rest of the paper is organized as follows. The most related work is investigated in Section \ref{sec:related}. Section \ref{sec:formulation} presents our problem formulation. The ZD alliance  strategy for a sequential multiple-player repeated game is deduced in Section \ref{sec:zd}, and our incentive algorithm is proposed in Section \ref{sec:op}. We report our simulation results on the proposed algorithm in Section \ref{sec:simu}, and Section \ref{sec:conclusion} concludes the paper.

\section{Related Work}
\label{sec:related}

Over the last two decades, research on cooperation
induction
has made considerable progress, which can be categorized into two types: reputation-based and pricing-based.

Reputation-based methods employ the concept of reputation or quasi-reputation to  evaluate the behaviors of a node based on  its contributions  to others, which is also a criterion  for  reacting to the service requests of this node.
For example, in  \cite{Lu2010Pi}, relay nodes would get good
reputation values for their cooperation, which can build other nodes' confidence on them thus helping forward their bundles.  The in-network realization of incentives was proposed in  \cite{uddin2010relics} to attach an explicit ranking to a node in light of its transit behaviors and translate the rank into message priority. Hu {\it et al.} \cite{hu2009budget} proposed a budget-based self-optimized incentive search protocol for unstructured P2P file sharing systems, motivating selfish nodes to earn more credits by providing services to others.
As a game theoretical incentive,  {\it Multicent} \cite{chen2015multicent} assigns credits for packet forwarding/storage in proportional to the priorities specified in the routing strategy.  The concept of {\it virtual credit} was adopted in \cite{Ning15}  to encourage selfish nodes to cooperate in data forwarding.

Pricing-based methods take advantage of monetary incentives to stimulate cooperation.  For instance,  Ning {\it et al.} \cite{ning13} introduced the concept of {\it virtual checks} to pay the cooperation of selfish nodes for ad dissemination in autonomous mobile social networks. Chen {\it et al.} \cite{chen} proposed an  auction-based incentive mechanism for paid content offloading considering the dual identity of service providers. Koutsopoulos \cite{Koutsopoulos} cast participatory sensing in the context of optimal reverse auction design, offering reasonable payments for  data contributors. Yang {\it et al.} \cite{Dejunmobicom} also proposed an   auction-based  mechanism to incentivize  participants of mobile phone sensing while allowing users to have more control over the payment they can receive.

Both reputation-based and pricing-based egoistic incentives aim to shape the behaviors of individuals and involve maintaining and managing the reputation or transaction states, leading to low efficiency in a large-scale system. In this paper, we take a dramatically different approach from the perspective of {\it a network} rather than {\it an individual} to achieve good scalability by employing a ZD strategy. ZD strategies \cite{press2012iterated} were firstly  proposed by Press and Dyson in
2012,  providing us a revolutionary understanding of simultaneous-move repeated  games. ZD strategies can enforce a fixed linear relationship between the expected payoffs of two players, indicating that a ZD player can control its  opponent's payoff  in a unilateral way. Since ZD  strategies were proposed, several studies have been carried out to enrich their theoretical hierarchy. For example,  the application of ZD strategies  was extended from a two-player simultaneous-move game to  a multi-player one in \cite{zhoutao}. The concept of {\it ZD alliance}  was first   proposed in \cite{pnus}, whose strategy was explored for  simultaneous-move multi-player games. In this paper, we employ a different theoretical approach to deduce the strategy of a ZD alliance for sequential multi-player games to serve our problem scenario.

\section{Problem Formulation}
\label{sec:formulation}

To analyze the behaviors of the players from a network perspective, we employ an indirect graph $G=(V,E)$ to describe a large-scale system, where  $V$ is the set of  nodes representing the players in the system and  $E$ is the set of edges denoting the social ties of the players. Each player has two choices in the game, namely {\it cooperation} $c$ and {\it defection} $d$. For example, in mobile opportunistic networks, $c$ means a player's willingness to transmit data for others and $d$ implies rejecting relay services.  Even though $c$ and $d$ have different meanings in different scenarios, they own common traits: the former is driven by social interests while the latter involves that of an individual. In practice, an application may involve multiple rounds and each player may choose $c$ or $d$ within a round; and some players may move first while the others take actions after observing the first movers. 
As a result, we focus on a sequential repeated game with multiple players in this paper.

In any game, a rational player aims to maximize its utility, which depends on not only the action of  itself but also those of others.  In this paper, we adopt a  method  similar to that in \cite{pnus,zhoutao} to calculate the utility $U_i(a_i|j)$ of a player $i$ when the number of cooperators among its neighbors is $j$: 
\begin{equation}
\label{eq:u}
U_i(a_i|j) = \frac{r(N_i)\big(j+a_i\big)}{N_i+1}+(1-a_i), ~a_i\in \{0,1\},
\end{equation}
where $a_i=1$ means player $i$ selects cooperation while $a_i=0$ implies defection; $N_i$ is the  number of neighbors  of player $i$, and obviously $0 \leq j\leq N_i$;  and $r(N_i)$ is the profit proportional to $N_i+1$. We use $r(N_i)=r$ in sequel for simplicity.

\begin{theorem}
\label{dilemma}
A social dilemma happens  when $r>1$.
\end{theorem}

\begin{IEEEproof}
According to \cite{pnas32}, a social  dilemma occurs when the utility of each player possesses the following properties: 1) a player, regardless its decision, can get a higher utility when more co-players select cooperation; 2) the utility of a defector is higher than that of  a cooperator; and 3) mutual cooperation outperforms mutual defection. As a result, the following inequalities should be satisfied, which are built based on the above three properties:
\begin{align}
\begin{cases}
  U_i(a_i|j+1) \geq U_i(a_i|j),~ 0 \leq j \leq N_i-1, \notag \\
  U_i(0|j+1)>U_i(1|j),~ 0 \leq j \leq N_i-1, \notag \\
  U_i(1|N_i)>U_i(0|0). \notag
\end{cases}
\end{align}

According to (\ref{eq:u}), the first two inequalities obviously hold without any constraint while for the last one, it is sufficient that $r>1$. Hence, the theorem is proved.
\end{IEEEproof}

According to Theorem \ref{dilemma}, a defector can obtain a higher utility than a cooperator. However, once all players adopt defection to act as {\it free riders}, each of which only gets the utility of 1 according to (\ref{eq:u}), which is lower than the utility of $r$ when all players cooperate because $r>1$, leading to the tragedy of the commons. To deter free riders, a natural idea is to teach players to get the cognition that  cooperation outperforms defection, synchronizing self-interest with the common one. This requires an egoistic incentive by an explicit side payment in the form of rewarding cooperation while punishing defection. However, it is inefficient to employ centralized methods  for the egoistic incentives especially in a large-scale system due to the lack of flexibility, robustness, and {\it the single point of failure} issue. Hence, in this paper, we resort to a distributed one.

\begin{figure}
\centering
\includegraphics[width=0.3\textwidth]{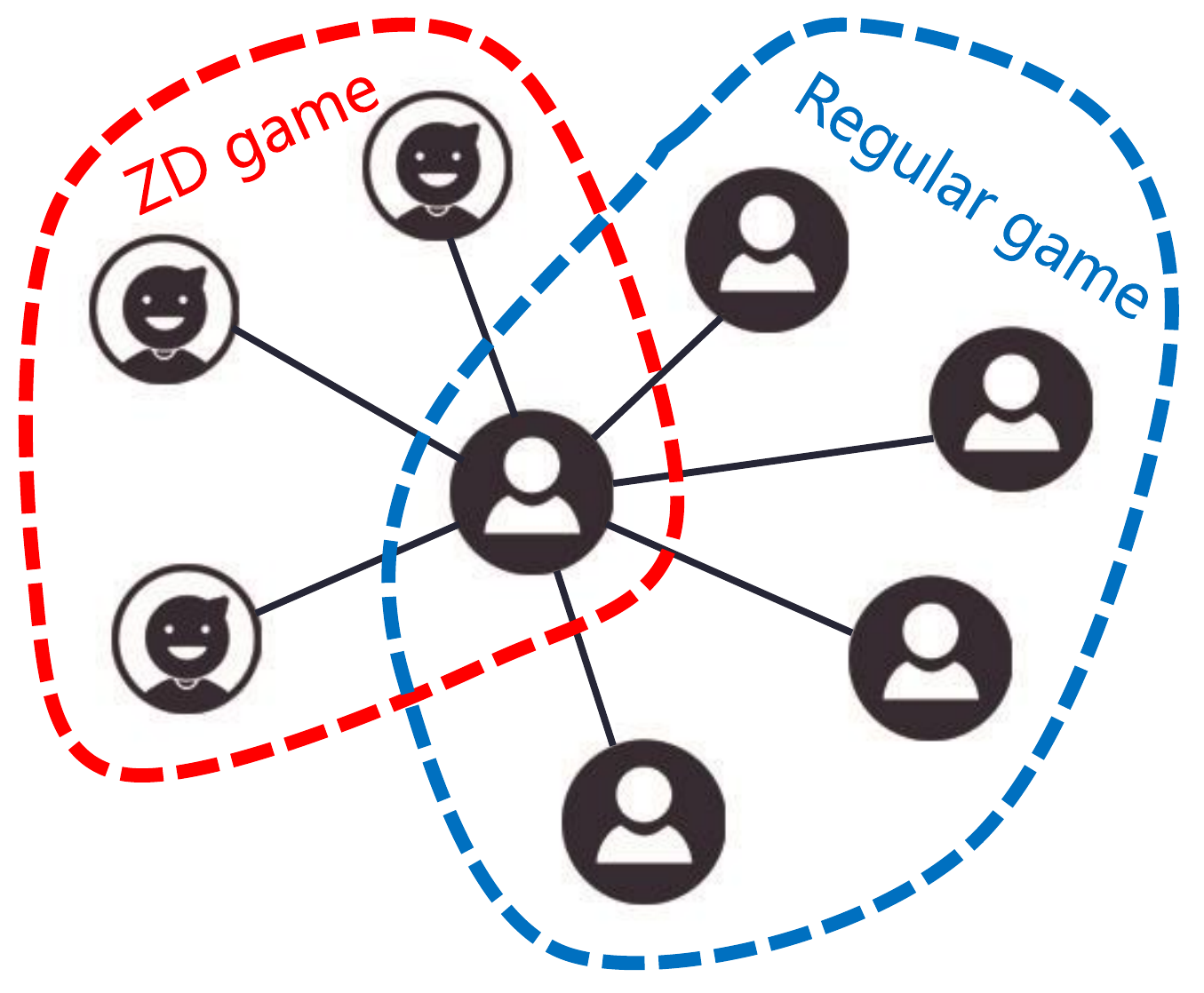}
\caption{Two types of games involving any regular player.}
\label{fig:tt}
\end{figure}

To realize a distributed egoistic incentive, we take advantage of ZD strategies \cite{press2012iterated}. A ZD player can set its opponent's payoff  irrespective of what action the opponent would take. Such capability  is very valuable for shaping the behavior of free riders. Hence, in this paper,  we address the social dilemma problem through the deployment of ZD players forming an alliance in the network.  We call  players not adopting the ZD strategy {\it regular players}. Then,  as shown in Fig. \ref{fig:tt},  in a network with heterogeneous players, the utility $\widetilde{U_i}$ of any regular player $i$  may consist of two components: the one obtained from the game playing with other regular players and the other from  playing with the ZD
alliance. The former game is called {\it a regular game} while the latter one is called {\it a ZD game}. Particularly,
\begin{equation}
\label{eq:newu}
\widetilde{U_i}(a_i|n_i,n^{\mathcal{A}}) = U_i(a_i|n_i)+ u_i(a_i|n^{\mathcal{A}}), \\~a_i\in \{0,1\}.
\end{equation}

In (\ref{eq:newu}),  $U_i(a_i|n_i)$ is the utility obtained from the regular game when player $i$ acts $a_i \in \{0,1\}$ playing with $n_i$ cooperators among the regular neighbors, which can be calculated by   (\ref{eq:u}); and $u_i(a_i|n^{\mathcal{A}})$ is the utility set by the ZD players when player $i$ acts $a_i \in \{0,1\}$ and there are  $n^{\mathcal{A}}$  ZD players as its neighbors.

Let the probability of a player choosing $c$ or $d$ be its strategy. Being rational, a player chooses a strategy according to the utility brought by cooperation or defection, which is also affected by the co-players' moves. Hence, we use the following  equation, similar to the one in \cite{zhao},  to determine the  strategy of player $i$ at each round:
 \begin{equation}
\label{eq:newstrategy}
q_i(c)  = \frac{e^{\widetilde{U_i}(1|n_i, n^{\mathcal{A}})-\widetilde{U_i}(0|n_i+1,n^{\mathcal{A}})}}{1+e^{\widetilde{U_i}(1|n_i,n^{\mathcal{A}})-\widetilde{U_i}(0|n_i+1,n^{\mathcal{A}})}}.
\end{equation}
In  (\ref{eq:newstrategy}), $q_i(c)$ is the probability that player $i$ cooperates. Obviously, the probability that player $i$ defects is $1-q_i(c)$. Here, $q_i(c)$ depends on the difference between $\widetilde{U_i}(1|n_i, n^{\mathcal{A}})$ and $\widetilde{U_i}(0|n_i+1,n^{\mathcal{A}})$, the utilities of player $i$ adopting $c$ and $d$, respectively, when the number of regular cooperators involving in the game is $n_i+1$ and there are $n^{\mathcal{A}}$ ZD players in the social community\footnote{$1|n_i$ means player $i$ adopts cooperation when  the number of other players' adopting $c$ is  $n_i$  and hence  the total number of cooperators is $n_i+1$ in this case.}.  
Note that both the number of regular cooperators and that of ZD players can be obtained according to historical information of games. 
As a result,  our optimization problem turns out to be the following: {\it how to deploy ZD players to maximize the overall cooperation probability in the whole system, where each regular player's cooperation probability is calculated by (\ref{eq:newstrategy})?}

\section{ZD alliance strategy in sequential multiple-player repeated game}
\label{sec:zd}

To solve our optimization problem, we need to determine $u_i(a_i|n^{\mathcal{A}})$ in (\ref{eq:newu}), the utility set by $n^{\mathcal{A}}$ ZD players when a regular player $i$ acts $a_i\in \{0,1\}$. Note that for any player $i$, there might exist more than one ZD players in its neighborhood. In this case, multiple ZD players must take the same action for the same move made by player $i$ because they should abide by consistent rules for being fair-minded regulators in our approach. The ZD players taking  the same action form a {\it ZD alliance}. In this section, we need to analyze how a ZD alliance sets the  utility of a regular player  to determine $u_i(a_i|n^{\mathcal{A}})$. Particularly, when $n^{\mathcal{A}}=1$, a ZD alliance strategy regresses to a classical ZD strategy. Note that, the original or variant  ZD strategies \cite{press2012iterated, zhoutao, pnus} are only  applicable to simultaneous-move repeated games with  two or more players, which are not suitable for our scenario. Hence,  we first extend the application domain of ZD strategies for a sequential multiple-player repeated game.

In a sequential multiple-player repeated game, players are classified into two types: the first-move players and the second-move ones. We call them {\it leaders} and {\it followers} respectively in this paper.  No matter which type a player belongs to, its utility is calculated according to (\ref{eq:u}), which is affected by the co-players' moves. Because a leader has to move first, its strategy is made based on all players' actions in the previous round.  Let $\mathbf{p}^{i}$ be the strategy of leader ${i}$. We have
\begin{equation}
\begin{aligned}
\label{p}
\mathbf{p}^{i}=(&p_{c,n^{\mathcal{L}}-1,N-n^{\mathcal{L}}}^{i},\cdots,p_{c,n^{\mathcal{L}}-1,0}^{i},\cdots,p_{c,0,0}^{i},\\
&p_{d,n^{\mathcal{L}}-1,N-n^{\mathcal{L}}}^{i},\cdots,p_{d,n^{\mathcal{L}}-1,0}^{i},\cdots,p_{d,0,0}^{i}),\\
&{i}\in \{1,2,\ldots, n^{\mathcal{L}}\},
\end{aligned}
\end{equation}
where each element can be represented in the form of $p^i_{s_i,x, y}$,  the probability of leader $i$ choosing $c$ given that it played $s_i \in \{c,d\}$  previously and the numbers of cooperators among other leaders  and followers were respectively $x$ and $y$ in the previous round; and $n^{\mathcal{L}}$ and $N$ are respectively the numbers of leaders and  all players in this game.

Since a follower can observe the behavior of the leaders in the current round, its strategy depends on the leaders' moves. Let $\mathbf{q}^{j}$ be the strategy of follower $j$, which can be represented as
\begin{equation}
\label{q}
\mathbf{q}^{j}=(q_{n^{\mathcal{L}}}^{j},q_{{n^{\mathcal{L}}}-1}^{j},\cdots,q_{0}^{j}), j \in \{n^{\mathcal{L}}+1,n^{\mathcal{L}}+2,\ldots, N\},
\end{equation}
where each element can be written in the form of $q_{z}^{j}$, the probability of follower $j$ choosing $c$ when there are $z$ cooperators  among the leaders.

With the  definitions of $\mathbf{p}^{i}$ and $\mathbf{q}^{j}$, $i \in \{1,2,\ldots, n^{\mathcal{L}}\}$ and $j \in \{n^{\mathcal{L}}+1,n^{\mathcal{L}}+2,\ldots, N\}$, we can construct the following  Markov matrix
\begin{equation}
\mathbf{M}=[M_{vw}]_{2^N\times2^N}, \nonumber
\end{equation}
where each element $M_{vw}$ denotes the one-step transition probability from state $v$ to  $w$. Here, a state denotes the moves made by all players in this round.  The state transition probability is essentially a joint one that can be obtained by
\begin{equation}
M_{vw}=\prod_{i=1}^{n^{\mathcal{L}}}\alpha_{i}\prod_{j=n^{\mathcal{L}}+1}^{N}\beta_{j}, \nonumber
\end{equation}
in which $\alpha_i$ and $\beta_j$ respectively denote the probabilities of actions made by leader $i$ and follower $j$  in state $w$ and they can be calculated as
\begin{equation}
\label{a}
\alpha_i=(p^i_{s_i,x,y})^{a_{i}}(1-p^i_{s_i,x,y})^{1-a_i}
\end{equation}
and
\begin{equation}
\label{b}
\beta_{j}=(q_{z}^{j})^{a_j}(1-q_{z}^{j})^{1-a_j}.
\end{equation}

In (\ref{a}), $p^i_{s_i,x,y}$ also denotes  the probability of leader $i$ cooperating when it  played $s_i \in \{c,d\}$ and the numbers of cooperators among other leaders  and followers were respectively $x$ and $y$ in the last round (state $v$);  and $a_i \in \{0,1\}$ is the action of leader $i$ made in state $w$. In (\ref{b}),  $q_{z}^{j}$ is also the cooperation probability of follower $j$  when there are $z$ cooperators  among the leaders
in state $w$ and  $a_j \in \{0,1\}$ is its action in state $w$.

Define a matrix $\mathbf{M'}=\mathbf{M}-\mathbf{I}$, where $\mathbf{I}$ is the unitary matrix.
Let $\mathbf{v}$ be the stable vector of the transition matrix $\mathbf{M}$; then we have $\mathbf{v}^\mathrm{T} \mathbf{M} = \mathbf{v}^\mathrm{T}$; hence $\mathbf{v}^\mathrm{T} \mathbf{M'} = 0$. Denote $[M_1, M_2, \ldots, M_m, \ldots, M_{2^N}]$ by $\mathbf{M'}$, where $M_i$ is the $i^{th}$  column vector.  In light of Cramer's rule, $Adj(\mathbf{M'})\mathbf{M'}=det(\mathbf{M'})\mathbf{I}=\mathbf 0$, where $Adj(\mathbf{M'})$ is the adjugate matrix of $\mathbf{M'}$. Combining this equation with  $\mathbf{v}^\mathrm{T} \mathbf{M'} = 0$,  we know that $\mathbf{v}$ is proportional to each row of $Adj(\mathbf{M'})$ \cite{pnus}. Therefore,  the dot product of the stable vector $\mathbf{v}$ and any vector $\mathbf{f}=(f_1,f_2,\cdots,f_{2^{N}-1},f_{2^{N}})$ is
\begin{equation}
\mathbf{v} \cdot \mathbf{f} = \text{det}~[M_1, M_2, \ldots, M_m, \ldots, M_{2^N-1},\mathbf{f}]. \nonumber
\end{equation}

Now we apply an elementary column transformation to $[M_1, M_2, \ldots, M_m, \ldots, M_{2^N-1},\mathbf{f}]$: find a column corresponding to a state where only one leader cooperates (i.e., the element of that column is in the form of $p^i_{s_i,x,y}\prod_{j\neq i}(1-p^j_{s_j,x,y})\prod_k(1-q_1^k),~ i\in  \{1,2,\ldots, n^{\mathcal{L}}\}$, $j\in  \{1,2,\ldots, n^{\mathcal{L}}\}$, $k \in  \{n^{\mathcal{L}}+1,n^{\mathcal{L}}+2,\ldots, N\}$) and a set of columns (denoted as $\Theta$)  corresponding to the states where player $i$ and at least one co-player cooperates (i.e., each element of those columns is represented as $p^i_{s_i,x,y}\prod_{j\neq i}\alpha_j\prod_k\beta_{k}$, $i\in  \{1,2,\ldots, n^{\mathcal{L}}\}$,  $j\in  \{1,2,\ldots, n^{\mathcal{L}}\}$, $k \in  \{n^{\mathcal{L}}+1,n^{\mathcal{L}}+2,\ldots, N\}$, $\exists j$, $a_j=1$ $\vee$ $\exists k$, $a_{k}=1$); then add all columns in $\Theta$ to the  first column we found to form a new column whose element is  $p^i_{s_i,x,y}\mathcal{T}-1$ if a diagonal element of $\mathbf{M'}$ is added to this entry and otherwise is $p^i_{s_i,x,y}\mathcal{T}$, where $\mathcal{T}=\prod_{j\neq i}(1-p^j_{s_j,x,y})\prod_k(1-q_1^k)+\sum_{\Theta}\prod_{j\neq i}\alpha_j\prod_k\beta_{k}$. Obviously, $\mathcal{T}=1$.

 After the elementary column transformation,
the matrix $[M_1, M_2, \ldots, M_m, \ldots, M_{2^N-1},\mathbf{f}]$ has a new form, i.e.,
 \begin{equation}
\label{vf}
[M_1, M_2, \ldots, M'_m, \ldots, M_{2^N-1},\mathbf{f}], \nonumber
\end{equation}
 where $M'_m$ is the new column formed according to the above method. More specifically,
\begin{equation}
\begin{aligned}
\label{juz}
&  ~[M_1, M_2, \ldots, M'_m, \ldots, M_{2^N-1},\mathbf{f}]\\
& =  \left[
\begin{matrix}
\cdots&\  p^i_{s_i,x,y}-1&\cdots&f_{1}\\
\cdots&\  p^i_{s_i,x,y}-1&\cdots&f_{2}\\
\ddots &\vdots&\ddots&\vdots\\
\cdots&\  p^i_{s_i,x,y}&\cdots&f_{2^{N}-1}\\
\cdots&\  p^i_{s_i,x,y}&\cdots&f_{2^N}
\end{matrix}
\right].
\end{aligned}
\end{equation}

It is worthy of  noting that the new column can be located at any place of the original $[M_1, M_2, \ldots, M_m, \ldots, M_{2^N-1},\mathbf{f}]$  indicating a state where only one leader cooperates. Here, we assume leader $i$ is such a player, and the corresponding column is located at the $m^{th}$ position.
Notably, the $m^{th}$ column in (\ref{juz}) is only related to the strategy of leader $i$, denoted as
\begin{equation}
\mathbf{\widetilde{p}}=(p^i_{s_i,x,y}-1,p^i_{s_i,x,y}-1,\cdots,p^i_{s_i,x,y},p^i_{s_i,x,y}). \nonumber
\end{equation}
If we let $\mathbf{\widetilde{p}}=\phi\mathbf{f}$, where $\phi\neq 0$ is a coefficient,  the $m^{th}$ column is proportional to the last one, resulting in
\begin{equation}
\label{vf}
\mathbf{v} \cdot \mathbf{f}=0.
\end{equation}

To set a suitable $\mathbf{f}$, we
divide the players in the game into two groups : {\it alliance members} ($\mathcal{A}$) and {\it non-alliance members} ($\mathcal{-A}$), with the former being a subset of the leaders taking the same strategy as leader $i$ while the latter including the rest of the leaders and all followers.
Let vectors $\mathbf{g^{\mathcal{A}}}=(g_{s, b}^{\mathcal{A}})$ and $\mathbf{g^{\mathcal{-A}}}=(g_{s,b}^{\mathcal{-A}})$ be respectively the average payoffs of alliance members and the non-alliance members under all possible outcomes, where each element  $g_{s,b}^{\mathcal{A}}$ ($g_{s,b}^{\mathcal{-A}}$) is the average payoff of a member from the set $\mathcal{A}$ ($\mathcal{-A}$) when the alliance  adopts $s \in \{c,d\}$ and there are $b$ cooperators in total. Denote by $n^{\mathcal{A}}$ the number of alliance members. According to (\ref{eq:u}), we have
\begin{align}
\begin{cases}
\label{g}
  g_{c,b}^{\mathcal{A}}=\frac{rb}{N},~ n^{\mathcal{A}} \leq b \leq N , \\
  g_{d,b}^{\mathcal{A}}=\frac{rb}{N}+1,~ 0 \leq b \leq N- n^{\mathcal{A}},  \\
 g_{c,b}^{\mathcal{-A}}=\frac{(b-n^{\mathcal{A}})\frac{rb}{N}+(N-b)(\frac{rb}{N}+1)}{N-n^{\mathcal{A}}},~ n^{\mathcal{A}}\leq b \leq N,  \\
 g_{d,b}^{\mathcal{-A}}=\frac{b\frac{rb}{N}+(N-n^{\mathcal{A}}-b)(\frac{rb}{N}+1)}{N-n^{\mathcal{A}}}, 0 \leq b \leq N- n^{\mathcal{A}}.
\end{cases}
\end{align}

In (\ref{g}), all alliance members have the same utility  because their behaviors are the same. This is different from the non-alliance members whose utilities depend on not only their actions but also the number of cooperators and defectors from both sets of  $\mathcal{A}$ and $\mathcal{-A}$.

Based on the above analysis, one can set
\begin{equation}
\mathbf{f}=\chi(\mathbf{g^{\mathcal{A}}}-l\cdot\mathbf{1})-(\mathbf{g^{\mathcal{-A}}}-l\cdot\mathbf{1}), \nonumber
\end{equation}
where $\chi$ and $l$ are coefficients, and $\mathbf{1}$ is a vector with each element being 1. Let $\mathbf{\widetilde{p}}=\phi\mathbf{f}=\phi[\chi(\mathbf{g^{\mathcal{A}}}-l\cdot\mathbf{1})-(\mathbf{g^{\mathcal{-A}}}-l\cdot\mathbf{1})]$. According to (\ref{vf}), we have
\begin{eqnarray} \label{eq:I}
\mathbf{v}\cdot \mathbf{f}&  = & \mathbf{v}\cdot \big(\chi(\mathbf{g^{\mathcal{A}}}-l\cdot\mathbf{1})-(\mathbf{g^{\mathcal{-A}}}-l\cdot\mathbf{1})\big) \nonumber \\
& = & \chi(\mathbf{\mathbf{v}\cdot g^{\mathcal{A}}}-l\cdot\mathbf{v}\cdot\mathbf{1})-(\mathbf{v}\cdot\mathbf{g^{\mathcal{-A}}}-l\cdot\mathbf{v}\cdot\mathbf{1}) \nonumber \\
&= & \chi(\pi^{\mathcal{A}} -l)-(\pi^{\mathcal{-A}}-l) =0.
\end{eqnarray}

In (\ref{eq:I}), $\pi^{\mathcal{A}}$ and $\pi^{\mathcal{-A}}$ are respectively the expected utilities of alliance members and  non-alliance members. Recall that  $\mathbf{v}$  is the stable vector of the transition matrix; hence $\mathbf{\mathbf{v}\cdot g^{\mathcal{A}}}=\pi^{\mathcal{A}}$, $\mathbf{\mathbf{v}\cdot g^{\mathcal{-A}}}=\pi^{\mathcal{-A}}$ and $\mathbf{v}\cdot\mathbf{1}=1$. Obviously, (\ref{eq:I}) shows that when leader $i$ sets its strategy to  $\mathbf{\widetilde{p}}$, it can enforce a linear relationship between $\pi^{\mathcal{A}}$ and  $\pi^{\mathcal{-A}}$, i.e.,
\begin{equation}
\label{l}
\pi^{\mathcal{-A}} = \chi\mathbf{\pi^{\mathcal{A}}}+(1-\chi)l,
\end{equation}
 which means that leader $i$ acts as a ZD player in this situation.
 Since  leader $i$ is a member of the alliance and all  alliance members should take the same action,  this alliance takes a ZD strategy and is termed a ZD alliance. We call the non-ZD alliance members  {\it outsiders}. Then, when $\chi=0$, the ZD alliance can set the  expected utility of the outsiders with $\mathbf{\pi^{\mathcal{-A}}}=l$.

\begin{theorem}
\label{avut}
When $\frac{r-1}{2r}N < n^{\mathcal{A}} \leq \frac{r-1}{r}N$, the expected utility of the outsiders satisfies $\frac{r(N-n^{\mathcal{A}})}{N} \leq \pi^{\mathcal{-A}} \leq
\frac{rn^{\mathcal{A}}}{N}+1$.
\end{theorem}
\begin{IEEEproof}
According to (\ref{l}), if $\chi\neq 1$, $l=\frac{\pi^{\mathcal{-A}}- \chi\mathbf{\pi^{\mathcal{A}}}}{1- \chi}$, which is proportional to $-\pi^{\mathcal{A}}$ and $\pi^{\mathcal{-A}}$. Because the ZD alliance moves first, we start by analyzing the range of $l$ from $\pi^{\mathcal{A}}$. Given that there are $b$ cooperators, according to the nature of social dilemma, we have
\begin{equation}
\label{pia}
g_{c,b}^{\mathcal{A}}=\frac{rb}{N} \leq \pi^{\mathcal{A}} \leq g_{d,b}^{\mathcal{A}}=\frac{rb}{N}+1.
\end{equation}

In other words, the expected utility of the ZD alliance is higher when it defects rather than cooperates.  When the ZD alliance defects, the average utility of the outsiders is $g_{d,b}^{\mathcal{-A}}=\frac{b\frac{rb}{N}+(N-n^{\mathcal{A}}-b)(\frac{rb}{N}+1)}{N-n^{\mathcal{A}}}$; otherwise, it is $g_{c,b}^{\mathcal{-A}}=\frac{(b-n^{\mathcal{A}})\frac{rb}{N}+(N-b)(\frac{rb}{N}+1)}{N-n^{\mathcal{A}}}$. Combining the range of $\pi^{\mathcal{A}}$ given in (\ref{pia}), we can deduce
\begin{equation}
\begin{aligned}
\max_{0\leq b \leq N-n^{\mathcal{A}}}&\{\frac{\vartheta \cdot b}{N(N-n^{\mathcal{A}})(1-\chi)}+1\}\\
& \leq l \leq
\min_{n^{\mathcal{A}}\leq b \leq N}\{\frac{\vartheta \cdot b+N^{2}}{N(N-n^{\mathcal{A}})(1-\chi)}\}, \nonumber
\end{aligned}
\end{equation}
where $\vartheta=r(N-n^{\mathcal{A}})(1-\chi)-N$. When $\frac{r-1}{2r}N < n^{\mathcal{A}} \leq \frac{r-1}{r}N$ and $\chi=0$, we can obtain the expected utility range of the outsiders.
\end{IEEEproof}

According to Theorem \ref{avut}, when $n^{\mathcal{A}}=1$, which implies that there is only one ZD alliance member,  the controllable range of $\pi^{\mathcal{-A}}$ shrinks to $[\frac{r(N-1)}{N}, \frac{r}{N}+1]$. This demonstrates that a ZD alliance can obtain higher regulation leverage compared to a single ZD player.

\newcounter{TempEqCnt}
\setcounter{TempEqCnt}{\value{equation}}
\setcounter{equation}{1}
\begin{figure*}[]
\begin{equation}
 \begin{minipage}[c]{0.5\textwidth}
 \centering
 \includegraphics[width=1.0\textwidth]{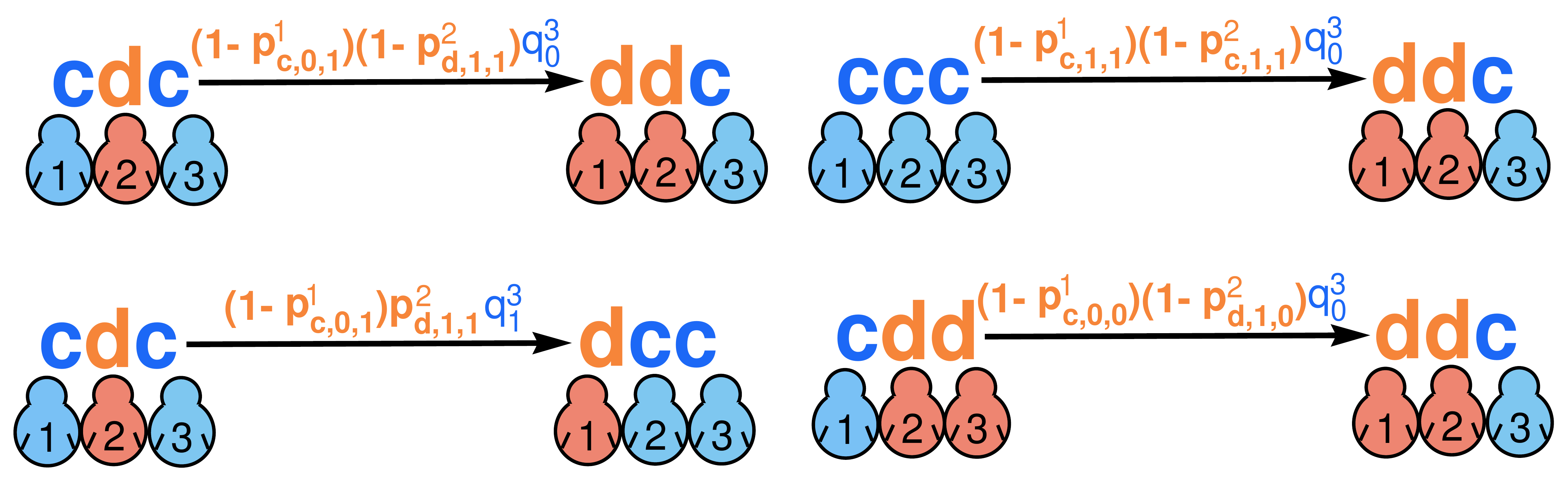}
 \end{minipage}
\begin{aligned}
\label{examp}
\ \ \ \ \ \ \text{Sta}&\text{te}:\{ccc,ccd,cdc,cdd,dcc,dcd,ddc,ddd\} \\
\ \ \ \ \ \ U_1&=[r, \frac{2r}{3}, \frac{2r}{3}, \frac{r}{3}, \frac{2r}{3}+1, \frac{r}{3}+1, \frac{r}{3}+1, 1 ]  \\
 U_2&=[r, \frac{2r}{3}, \frac{2r}{3}+1, \frac{r}{3}+1, \frac{2r}{3}, \frac{r}{3}, \frac{r}{3}+1, 1 ]  \\
 U_3&=[r, \frac{2r}{3}+1, \frac{2r}{3}, \frac{r}{3}+1, \frac{2r}{3}, \frac{r}{3}+1, \frac{r}{3}, 1 ]  \\
\mathbf{p}^{1}&=[p_{c,1,1}^{1},p_{c,1,0}^{1},p_{c,0,1}^{1},p_{c,0,0}^{1},p_{d,1,1}^{1},p_{d,1,0}^{1},p_{d,0,1}^{1},p_{d,0,0}^{1}]  \\
\mathbf{p}^{2}&=[p_{c,1,1}^{2},p_{c,1,0}^{2},p_{d,1,1}^{2},p_{d,1,0}^{2},p_{c,0,1}^{2},p_{c,0,0}^{2},p_{d,0,1}^{2},p_{d,0,0}^{2}]  \\
\mathbf{q}^{3}&=[q_{2}^{3},q_{1}^{3},q_{0}^{3}] \nonumber
\hfill
\end{aligned}
\end{equation}
\begin{tiny} 
\begin{equation}
\begin{aligned}
\label{examp}
&\mathbf{v} \cdot \mathbf{f} =D(\mathbf{p}^{1},\mathbf{p}^{2},\mathbf{q}^{3}, \mathbf{f}) =  \\
&{\text det} \left[
\begin{matrix}
p_{c,1,1}^{1}p_{c,1,1}^{2}q_{2}^{3}-1   & p_{c,1,1}^{1}p_{c,1,1}^{2}(1-q_{2}^{3}) & p_{c,1,1}^{1}(1-p_{c,1,1}^{2})q_{1}^{3}      &  \textcolor[rgb]{1.00,0.00,0.00}{p_{c,1,1}^{1}-1} & (1-p_{c,1,1}^{1})p_{c,1,1}^{2}q_{1}^{3} & \textcolor[rgb]{0.00,0.50,1.00}{p_{c,1,1}^{2}-1}               & (1-p_{c,1,1}^{1})(1-p_{c,1,1}^{2})q_{0}^{3}  &  f_{1} \\
p_{c,1,0}^{1}p_{c,1,0}^{2}q_{2}^{3}     & p_{c,1,0}^{1}p_{c,1,0}^{2}(1-q_{2}^{3})-1   & p_{c,1,0}^{1}(1-p_{c,1,0}^{2})q_{1}^{3}      &  \textcolor[rgb]{1.00,0.00,0.00}{p_{c,1,0}^{1}-1} & (1-p_{c,1,0}^{1})p_{c,1,0}^{2}q_{1}^{3} & \textcolor[rgb]{0.00,0.50,1.00}{p_{c,1,0}^{2}-1}               &(1-p_{c,1,0}^{1})(1-p_{c,1,0}^{2})q_{0}^{3}  &  f_{2} \\
p_{c,0,1}^{1}p_{d,1,1}^{2}q_{2}^{3}     & p_{c,0,1}^{1}p_{d,1,1}^{2}(1-q_{2}^{3})     &p_{c,0,1}^{1}(1-p_{d,1,1}^{2})q_{1}^{3}-1      &  \textcolor[rgb]{1.00,0.00,0.00}{p_{c,0,1}^{1}-1} & (1-p_{c,0,1}^{1})p_{d,1,1}^{2} q_{1}^{3} & \textcolor[rgb]{0.00,0.50,1.00}{p_{d,1,1}^{2}}                  & (1-p_{c,0,1}^{1})(1-p_{d,1,1}^{2})q_{0}^{3}  &  f_{3} \\
p_{c,0,0}^{1}p_{d,1,0}^{2}q_{2}^{3}     & p_{c,0,0}^{1}p_{d,1,0}^{2}(1-q_{2}^{3})     & p_{c,0,0}^{1}(1-p_{d,1,0}^{2})q_{1}^{3}      &  \textcolor[rgb]{1.00,0.00,0.00}{p_{c,0,0}^{1}-1}& (1-p_{c,0,0}^{1})p_{d,1,0}^{2}q_{1}^{3} & \textcolor[rgb]{0.00,0.50,1.00}{p_{d,1,0}^{2}}                  & (1-p_{c,0,0}^{1})(1-p_{d,1,0}^{2})q_{0}^{3}  &  f_{4} \\
p_{d,1,1}^{1}p_{c,0,1}^{2}q_{2}^{3}     & p_{d,1,1}^{1}p_{c,0,1}^{2}(1-q_{2}^{3})    & p_{d,1,1}^{1}(1-p_{c,0,1}^{2})q_{1}^{3}      &  \textcolor[rgb]{1.00,0.00,0.00}{p_{d,1,1}^{1}} & (1-p_{d,1,1}^{1})p_{c,0,1}^{2}q_{1}^{3}-1 & \textcolor[rgb]{0.00,0.50,1.00}{p_{c,0,1}^{2}-1}                 & (1-p_{d,1,1}^{1})(1-p_{c,0,1}^{2})q_{0}^{3}  &  f_{5} \\
p_{d,1,0}^{1}p_{c,0,0}^{2}q_{2}^{3}     & p_{d,1,0}^{1}p_{c,0,0}^{2}(1-q_{2}^{3})     & p_{d,1,0}^{1}(1-p_{c,0,0}^{2})q_{1}^{3}      &  \textcolor[rgb]{1.00,0.00,0.00}{p_{d,1,0}^{1}} & (1-p_{d,1,0}^{1})p_{c,0,0}^{2}q_{1}^{3} & \textcolor[rgb]{0.00,0.50,1.00}{p_{c,0,0}^{2}-1}                 & (1-p_{d,1,0}^{1})(1-p_{c,0,0}^{2})q_{0}^{3}  &  f_{6} \\
p_{d,0,1}^{1}p_{d,0,1}^{2}q_{2}^{3}     & p_{d,0,1}^{1}p_{d,0,1}^{2}(1-q_{2}^{3})     & p_{d,0,1}^{1}(1-p_{d,0,1}^{2})q_{1}^{3}      &  \textcolor[rgb]{1.00,0.00,0.00}{p_{d,0,1}^{1}} & (1-p_{d,0,1}^{1})p_{d,0,1}^{2}q_{1}^{3} & \textcolor[rgb]{0.00,0.50,1.00}{p_{d,0,1}^{2}}                 & (1-p_{d,0,1}^{1})(1-p_{d,0,1}^{2})q_{0}^{3}-1  &  f_{7} \\
p_{d,0,0}^{1}p_{d,0,0}^{2}q_{2}^{3}     & p_{d,0,0}^{1}p_{d,0,0}^{2}(1-q_{2}^{3})     & p_{d,0,0}^{1}(1-p_{d,0,0}^{2})q_{1}^{3}      & \textcolor[rgb]{1.00,0.00,0.00}{p_{d,0,0}^{1} }& (1-p_{d,0,0}^{1})p_{d,0,0}^{2}q_{1}^{3} & \textcolor[rgb]{0.00,0.50,1.00}{p_{d,0,0}^{2}}                  & (1-p_{d,0,0}^{1})(1-p_{d,0,0}^{2})q_{0}^{3}  &  f_{8} \nonumber
\end{matrix}
\right]
\end{aligned}
\end{equation}
\end{tiny}
\caption{An example sequential three-player repeated game.}
\end{figure*}
\setcounter{equation}{\value{TempEqCnt}}

We use  an example shown in Fig. \ref{examp} to explain key terms and definitions involved in this section. Assume that there are three players in the game, among which players 1 and 2 are the leaders while player 3 is the follower. The states of the game include all  possible outcomes, namely
\begin{equation}
\{ccc, ccd, cdc, cdd, dcc, dcd, ddc, ddd\} \nonumber.
\end{equation}
Accordingly, the utility of any player in each outcome can be calculated  according to (\ref{eq:u}).

The strategy of each leader and the follower  can be represented in light of (\ref{p}) and (\ref{q}).  For instance, given a previous outcome $cdc$, the conditional probability under which player 1 adopts $c$ in the current round is $p_{c,0,1}^{1}$ because it acted $c$ and there was no other cooperative leader but one cooperative follower in the previous round; similarly, player 2 adopts $c$ with the conditional probability of $p_{d,1,1}^{2}$ in the current round since it acted $d$ and both the remaining leader and the follower cooperated in the previous round; player 3 is a follower, whose strategy space's cardinality reduces to $n^\mathcal{L}+1$, which is much lower than that of any leader (i.e., $2^N$). The  conditional probability of player 3 adopting $c$ is $q_{z}^{3}$,  depending on the number ($z$) of cooperators among the leaders in the current round.   As a result, the probability from state  $cdc$ to any other state can be derived. For example, the transition probability to state $ddc$ is $(1-p_{c,0,1}^{1})(1-p_{d,1,1}^{2})q_{0}^{3}$. All state transition probabilities  contribute to the transition matrix $\mathbf{M}$. Let $\mathbf{M'}=\mathbf{M}-\mathbf{I}$, which can be denoted as
 \begin{equation}
[M_1, M_2, M_3, M_4, M_5, M_6, M_7, M_8]. \nonumber
\end{equation}

Next, we perform an elementary column  transformation on matrix $[M_1, M_2, M_3, M_4, M_5, M_6, M_7, \mathbf{f}]$, identifying a column corresponding to a state where only one leader cooperates (e.g., player 1) and adding to it all the other columns whose states indicate that player 1 cooperates and at least one of players 2 and 3 also selects $c$. The determinant after the elementary column  transformation becomes
   \begin{equation}
   [M_1, M_2, M_3, M_4', M_5, M_6, M_7, \mathbf{f}], \nonumber
\end{equation}
  where the fourth column, denoted  by the red color in Fig. \ref{examp}, is formed according to the above method. We can find that $M_4'$ is only dependent on the strategy of player 1. If we apply an elementary column transformation corresponding to player 2, we can obtain the sixth column (denoted by the blue color in Fig. \ref{examp}) that only relies on player 2.

 Let player 1 be the ZD player and ally with player 2. Both can set their strategies as $\mathbf{p}^{1}=\mathbf{p}^{2}=\phi[\chi(\mathbf{g^{\mathcal{A}}}-l\cdot\mathbf{1})-(\mathbf{g^{\mathcal{-A}}}-l\cdot\mathbf{1})]$, where $\mathbf{g^{\mathcal{A}}}= [g_{c,2}^{\mathcal{A}}, g_{c,3}^{\mathcal{A}}, g_{d,0}^{\mathcal{A}},g_{d,1}^{\mathcal{A}}]=[\frac{2r}{3}, r, 1, \frac{r}{3}+1]$  and $\mathbf{g^{\mathcal{-A}}}= [g_{c,2}^{\mathcal{-A}}, g_{c,3}^{\mathcal{-A}}, g_{d,0}^{\mathcal{-A}},g_{d,1}^{\mathcal{-A}}]=[\frac{2r}{3}+1, r, 1, \frac{r}{3}]$.    According to Theorem   \ref{avut}, the ZD alliance including players 1 and 2 can set the expected utility of player 3 ranging from $\frac{r}{3}$ to $\frac{2r}{3}+1$.

 Now, it is the time to answer the question proposed at the beginning of this section. Taking advantage of Theorem \ref{avut}, we can set
\begin{align}
\label{ui}
u_i(a_i|n^{\mathcal{A}})=
\begin{cases}
\frac{rn^{\mathcal{A}}}{N}+1,~ a_i=1,  \\
  \frac{r(N-n^{\mathcal{A}})}{N},~ a_i=0,
\end{cases}
\end{align}
which means a ZD alliance would reward the cooperation of the outsiders with the highest expected utility while punish their defection with the lowest utility. This extreme reward-punishment incentive mechanism can help to  facilitate cooperation as much as possible.

\section{ZD alliance for egoistic incentive}
\label{sec:op}

Based on our analysis in Section \ref{sec:formulation}, one can see that a rational player $i$ would take an action according to
the difference between the utilities of adopting $c$ and $d$, which are related to the co-players' moves among its social community. Due to the heterogeneous nature of the members in the social community, the utility of player $i$ may come from the regular game and the  ZD game, with the former being calculated by (\ref{eq:u}) while the latter solved by (\ref{ui}). Note that Theorem \ref{avut} in  Section \ref{sec:zd} can serve for a more general case,   where the number of outsiders is arbitrary. However, in our consideration, for any player $i$,  the corresponding ZD game only involves  the ZD alliance and the player itself as shown in Fig. \ref{fig:tt}, implying that the outsider is just  player $i$ and hence we have $N-n^{\mathcal{A}}=1$.

    From the perspective of a single player, one can realize egoistic incentives by rewarding more expected utility when the player chooses $c$ and punishing it if it defects. However, to achieve this goal from the network perspective, we need to optimize the deployment of ZD players so that the overall cooperation probability in the whole system can be maximized.

To optimize the above problem, we need to answer such a question: according to (\ref{l}), $\frac{\pi^{\mathcal{A}}-l}{\pi^{\mathcal{-A}}-l}=\frac{1}{\chi}$, implying that a ZD alliance has the ability to carry out the extortionate strategy by enforcing a ratio between its expected utility and that of the outsider. Moreover, the smaller the $\chi$, the more the expected utility that can be transferred to the ZD alliance. In this case,   as a dominant player, can the ZD alliance push the outsider  to cooperate while it intends to defect for obtaining more expected utility?  We use the following theorem to answer this question:
\begin{theorem}
\label{domi}
The dominant strategy of the ZD alliance is cooperation irrespective of the strategy adopted by the outsider.
\end{theorem}
\begin{IEEEproof}
In the ZD game, when player $i$ takes $a_i \in \{0,1\}$, the utilities of the ZD alliance when it cooperates and defects are respectively  $\frac{r(n^{\mathcal{A}}+a_i)}{N}$ and $\frac{ra_i}{N}+1$ according to (\ref{g}). To prove cooperation is the dominant strategy of the ZD alliance, we should prove $\frac{r(n^{\mathcal{A}}+a_i)}{N}>\frac{ra_i}{N}+1$, which is equivalent to proving $(r-1)n^{\mathcal{A}}>1$ due to $N-n^{\mathcal{A}}=1$ in our scenario.

In light of Theorem \ref{avut}, the condition under which a ZD alliance can control others is $\frac{r-1}{2r}N < n^{\mathcal{A}} \leq \frac{r-1}{r}N$, which can be simplified as $r>N$ due to $N-n^{\mathcal{A}}=1$. Because $N \geq 2$, we have  $(r-1)n^{\mathcal{A}}>1$. As a conclusion, this theorem holds.
\end{IEEEproof}

When cooperation is the dominant strategy of the ZD alliance,  maximizing the overall cooperation probability of the whole system is equivalent to maximizing those of all regular players. Hence, our optimization problem can be written as:
\begin{equation}
\begin{aligned}
\label{max}
&\max \ \sum_{i=1}^V q_i(c)x_i,\\
&s.t. \ \sum x_i=V-K, \\
    &  \ \ x_i \in \{0,1\},
\end{aligned}
\end{equation}
where $x_i=0$ denotes that player $i$ adopts  the ZD strategy and $x_i=1$ means that the player is a regular one; and $K$ and $V$ are respectively the number of ZD players we deploy and the number of total players in the system.

To find an optimal solution for (\ref{max}), we  resort to genetic algorithms (GA) because GA is efficient in dealing with problems with nonlinear and multi-constraint properties. As GA has been well-studied, we refer the interested readers to \cite{Mahmud2012Genetic} and omit the algorithm description here.  

The basic idea of our egoistic incentive mechanism is to take advantage of the ZD alliance to shape the behaviors of the regular players. According to (\ref{eq:newstrategy}), the cooperation probability of a regular player $i$ depends on the number of ZD players ($n^{\mathcal{A}}$) and that of cooperators ($n_i$) among the regular neighbors in its social community. However,  due to the nature of social dilemma \cite{pnas32}, we can obtain $U_i(1|n_i)-U_i(0|n_i+1)=\frac{r(n_i+1)}{\widetilde{N}}-(\frac{r(n_i+1)}{\widetilde{N}}+1)=-1$, where $\widetilde{N}$ is the total number of regular players in the social community. This equation reflects that a player $i$ can definitely  obtain a higher payoff if it defects rather than cooperates, no matter  how many cooperators there exist (i.e., $n_i$) in its neighborhood. This makes the strategy of a regular player $i$ only relies on $n^{\mathcal{A}}$, the number of ZD players around $i$. While $n^{\mathcal{A}}$ is determined by the solution of the optimization problem (\ref{max}), which is closely related to the topology of network $G$. Because $G$ is constructed based on the nodes' social ties, which are relatively steady, $n^{\mathcal{A}}$ is steady. Hence, our method is stable, and does not need to maintain and manage any state; thus achieving the property of statelessness introduced in Section \ref{sec:intro}.

\section{Performance Simulation}
 \label{sec:simu}

In this section,  we evaluate the performance of our proposed mechanism using real and synthetic data.

The real data we adopt is the {\it iMote data} \cite{cambridge-haggle}, including traces of Bluetooth sightings by groups of users carrying small devices (iMotes) for a number of days in different offices, conferences, and cities. We use all iMote nodes (54 in total) and their links to perform our simulation.
Synthetic data is also employed since we want to test the performance of our approach under different network topologies.  The types of networks in our simulation  are  {\it star}, {\it ring}, {\it tree}, and {\it mesh}, each of which includes 80 nodes. More specifically, all nodes  in the star network are individually connected to a central node; in the ring network,  each node connects to exactly two other nodes, forming a  closed loop; each node has two neighbors except the leaf nodes in the  tree network, making its shape a binary tree; and in the mesh network,  each node has at least two randomly selected neighbors. All simulations have been repeated 30 times to obtain the average values with statistical confidence.

\begin{figure}
\centering
\includegraphics[width=0.48\textwidth]{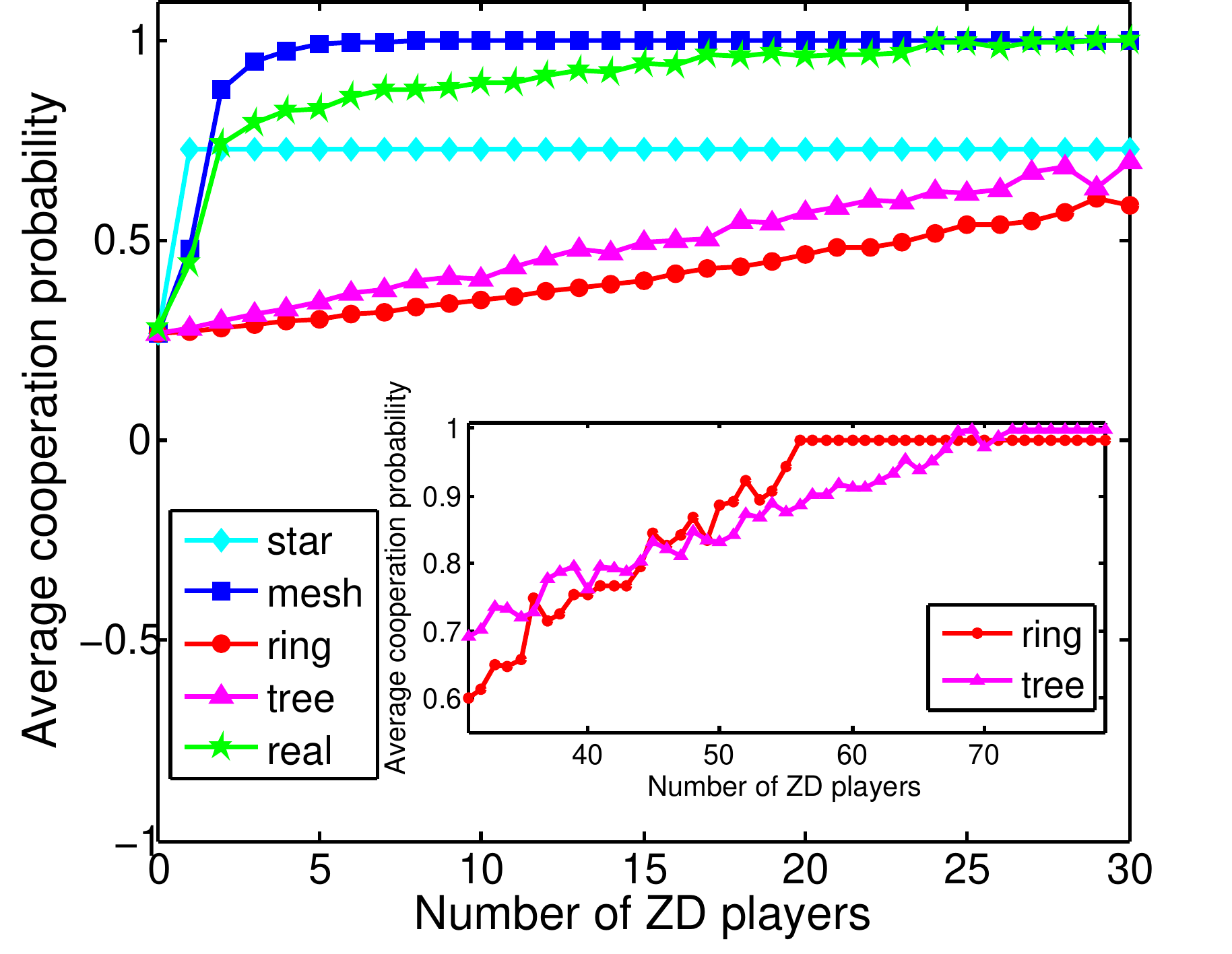}
\caption{Average cooperation probabilities in different networks when $r=2n+3$.}
\label{fig:total}
\end{figure}

Fig. \ref{fig:total} demonstrates how the average cooperation probability of our mechanism changes with the number of ZD players ($K$) deployed in different networks. 
In this study,  we set $r=2n+3$, where $n$ is the total number of players in a game. Note that each regular player may involve in two kinds of games, the regular game and the ZD game as shown in Fig. \ref{fig:tt}. Hence, $n$ varies in different games. 
According to Fig. \ref{fig:total}, one can see that the mesh network has the best performance, realizing fast cooperation over the whole system using fewer ZD players. Although the performance of the  real network  is slightly lower than that of the  mesh network, both have similar trends. This is because compared to other networks, the nodes in these two networks have higher degrees; thus fewer ZD players are needed to control more regular players.

Fig. \ref{fig:total} also shows that in the star network, once a ZD player is deployed, the average cooperation probability can rocket to  $73\%$ and keep this level even though the number of ZD players increases. The reason behind this lies in that each node has only one neighbor except the central node in  the star network, and a single ZD player is definitely needed in the central position to control all the other nodes while other ZD players completely lose their control power due to the lack of connections to other regular players. However, because of  only one ZD player functions, the controllable range of the outsider's expected utility shrinks to $[\frac{r}{2}, \frac{r}{2}+1]$, making the cooperation probability to be $\frac{e}{1+e}\approx 73\%$. This is different from the situations in the mesh network and the real network, where ZD players can achieve higher control leverage by forming alliances and finally obtain the desired cooperation probabilities.

Finally, one can see from Fig. \ref{fig:total} that the growth rates of cooperation probabilities in the tree and the ring networks  are the slowest. This is because the average degrees in the tree  and ring networks are respectively 1.95 and 2, which are about $\frac{1}{20}$ and  $\frac{1}{10}$  of those in the mesh and the real networks. Although the tree and ring networks have almost the same average degree as the star network, the ZD player located at the central place in the star network has the largest degree value, empowering its control.

\begin{figure}
\centering
\includegraphics[width=0.48\textwidth]{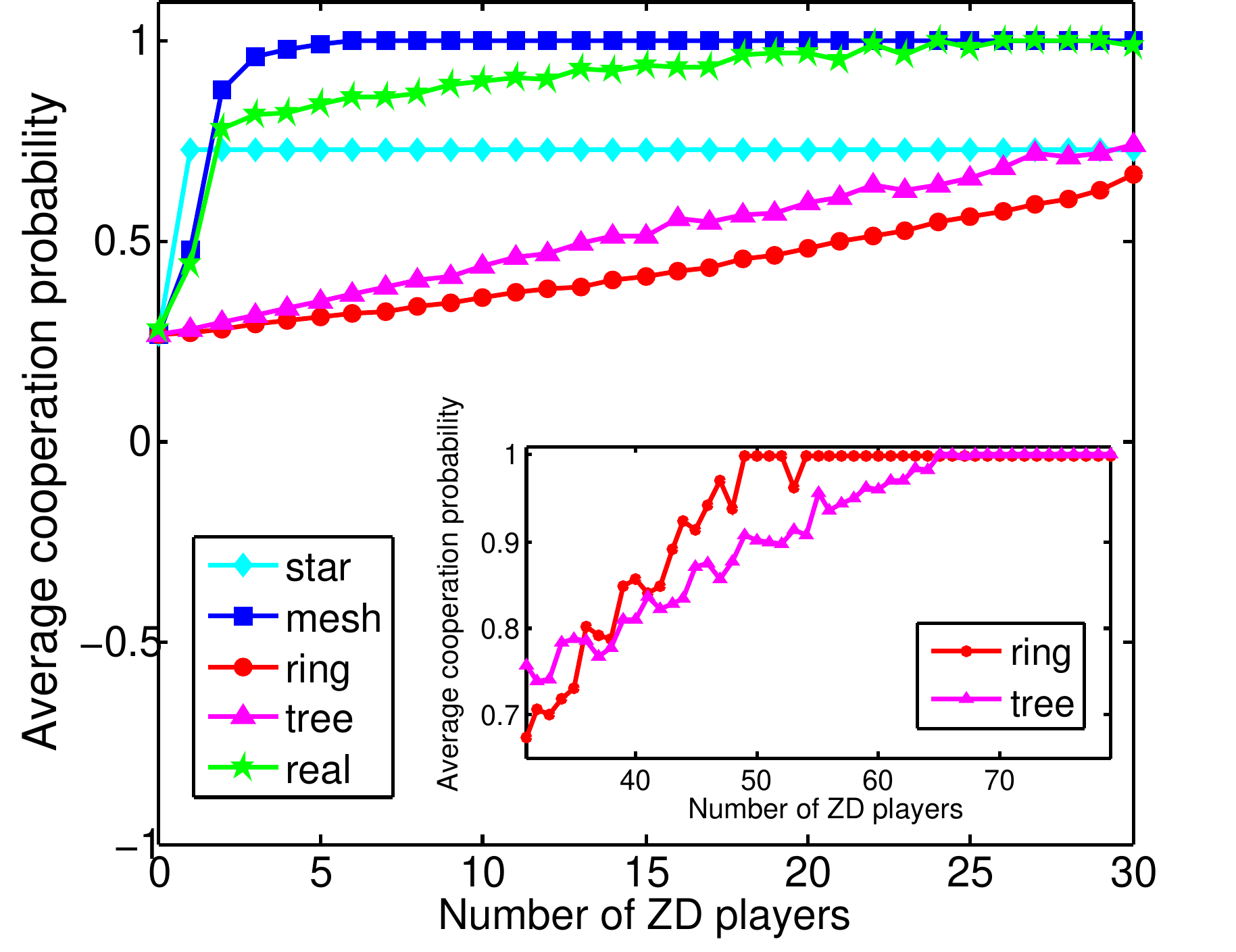}
\caption{Average cooperation probability in different networks when $r=2n^2+3$.}
\label{fig:total1}
\end{figure}

Fig. \ref{fig:total1} illustrates the average cooperation probabilities of our method under different numbers of ZD players ($K$) deployed  in different networks when $r=2n^2+3$, which implies a nonlinear relationship between $r$ and $n$. According to this figure, one can see that the results are similar to those shown in  Fig. \ref{fig:total1}. Note that we have done extensive simulations under other forms of $r(n)$ and obtained results with very similar trends; hence the corresponding results are omitted in this paper to avoid redundancy.

\begin{figure}[]
\centering
\subfloat[Cooperator ratio]{\includegraphics[width=1.9in,height=1.6in]{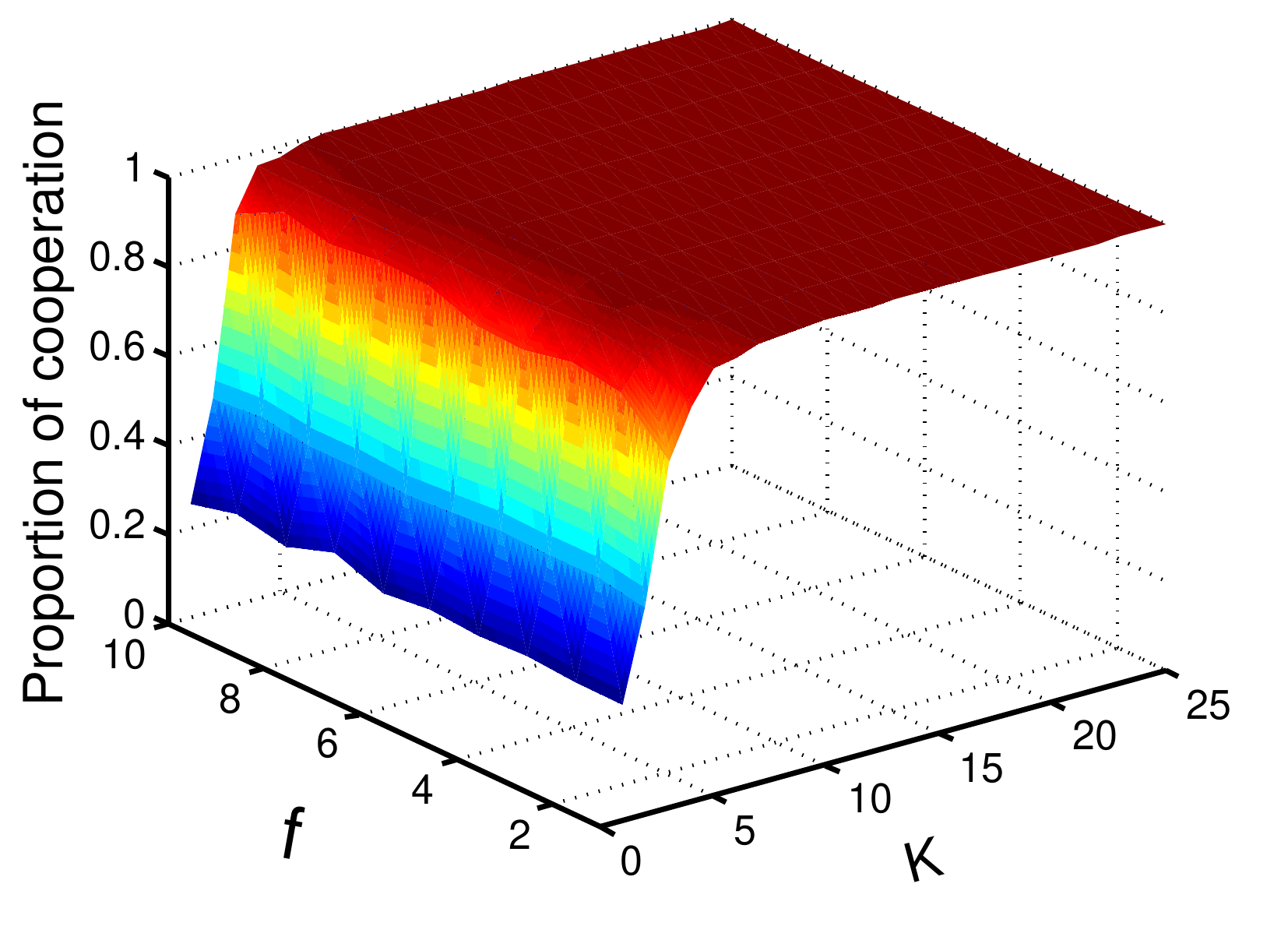}}
\subfloat[ZD deployment]{\includegraphics[width=1.4in,height=1.9in]{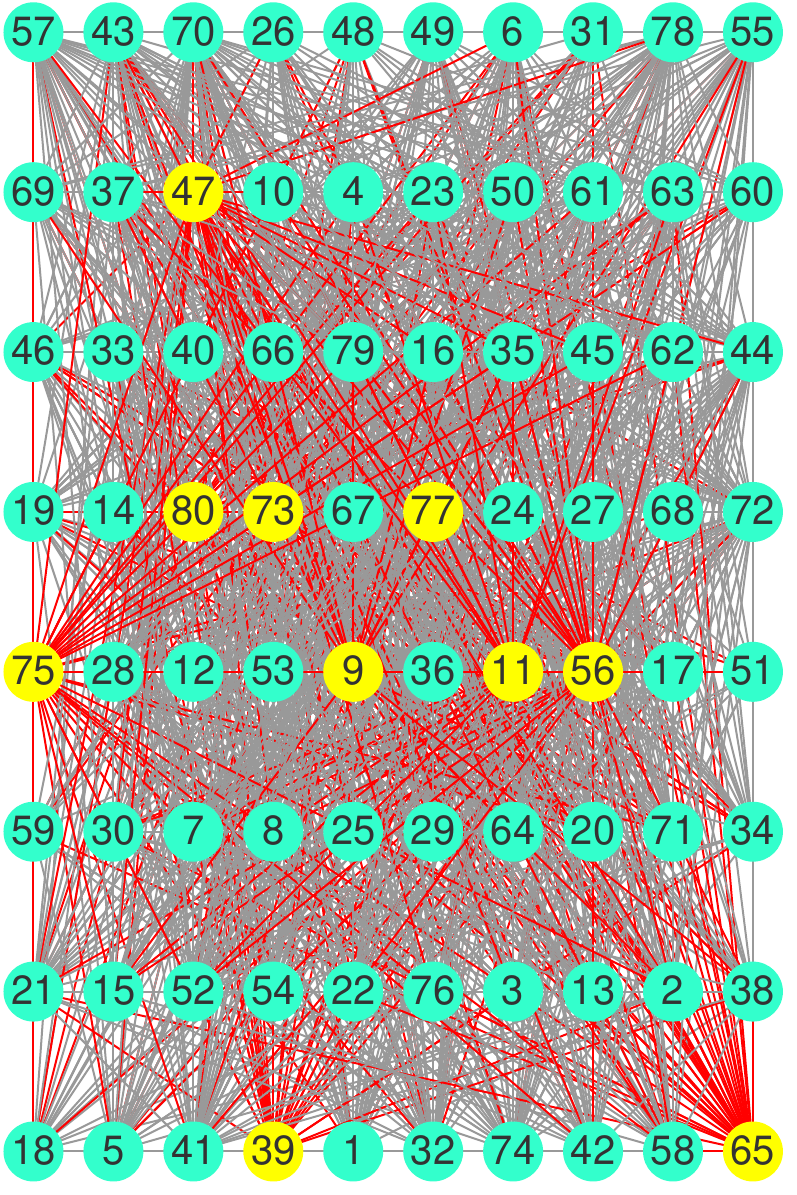}}
\hfill
\caption{Performance in the mesh network.}
\label{fig:mesh}
\end{figure}

\begin{figure}[]
\centering
\subfloat[Cooperator ratio]{\includegraphics[width=1.9in,height=1.6in]{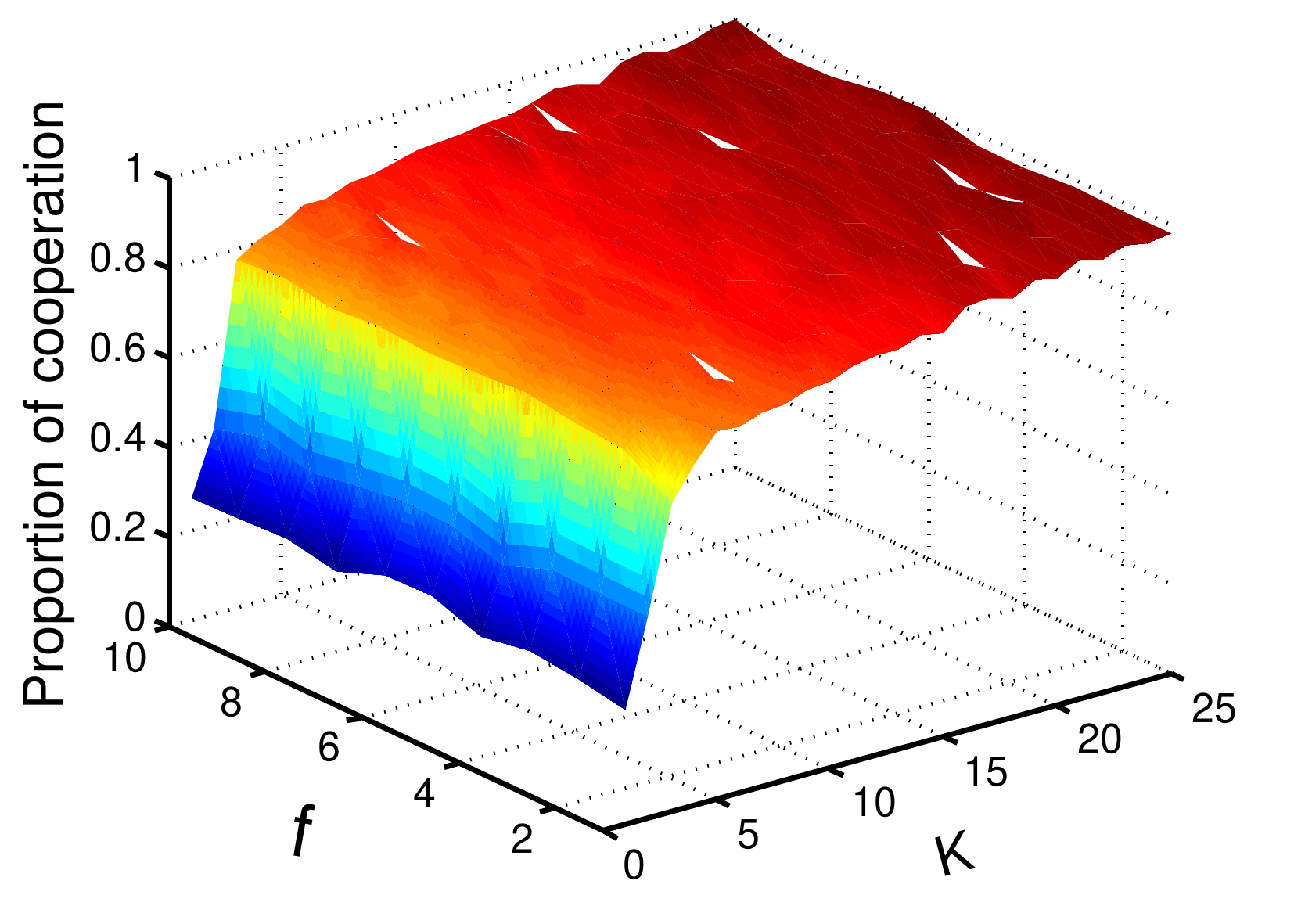}}
\subfloat[ZD deployment]{\includegraphics[width=1.4in,height=1.9in]{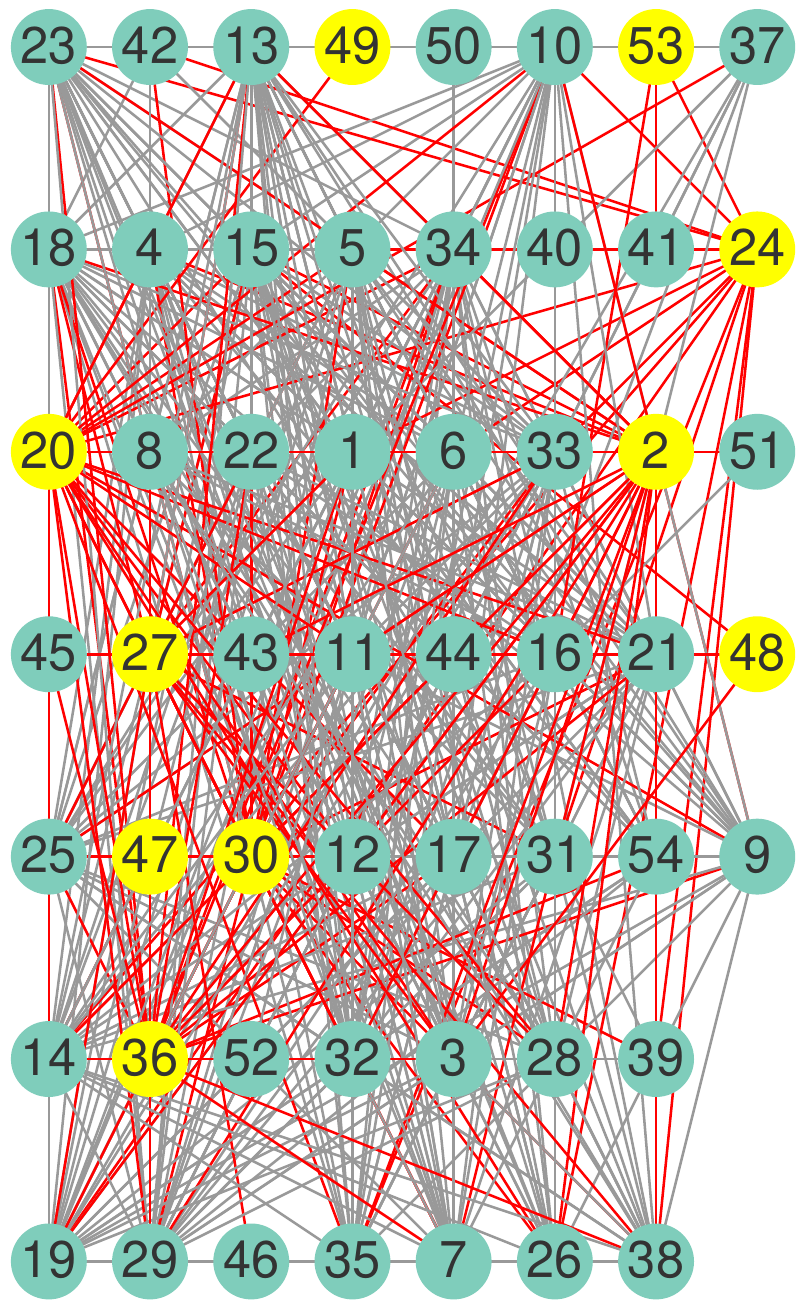}}
\hfill
\caption{Performance in the real  network.}
\label{fig:real}
\end{figure}

\begin{figure}[]
\centering
\subfloat[Cooperator ratio]{\includegraphics[width=1.7in,height=1.6in]{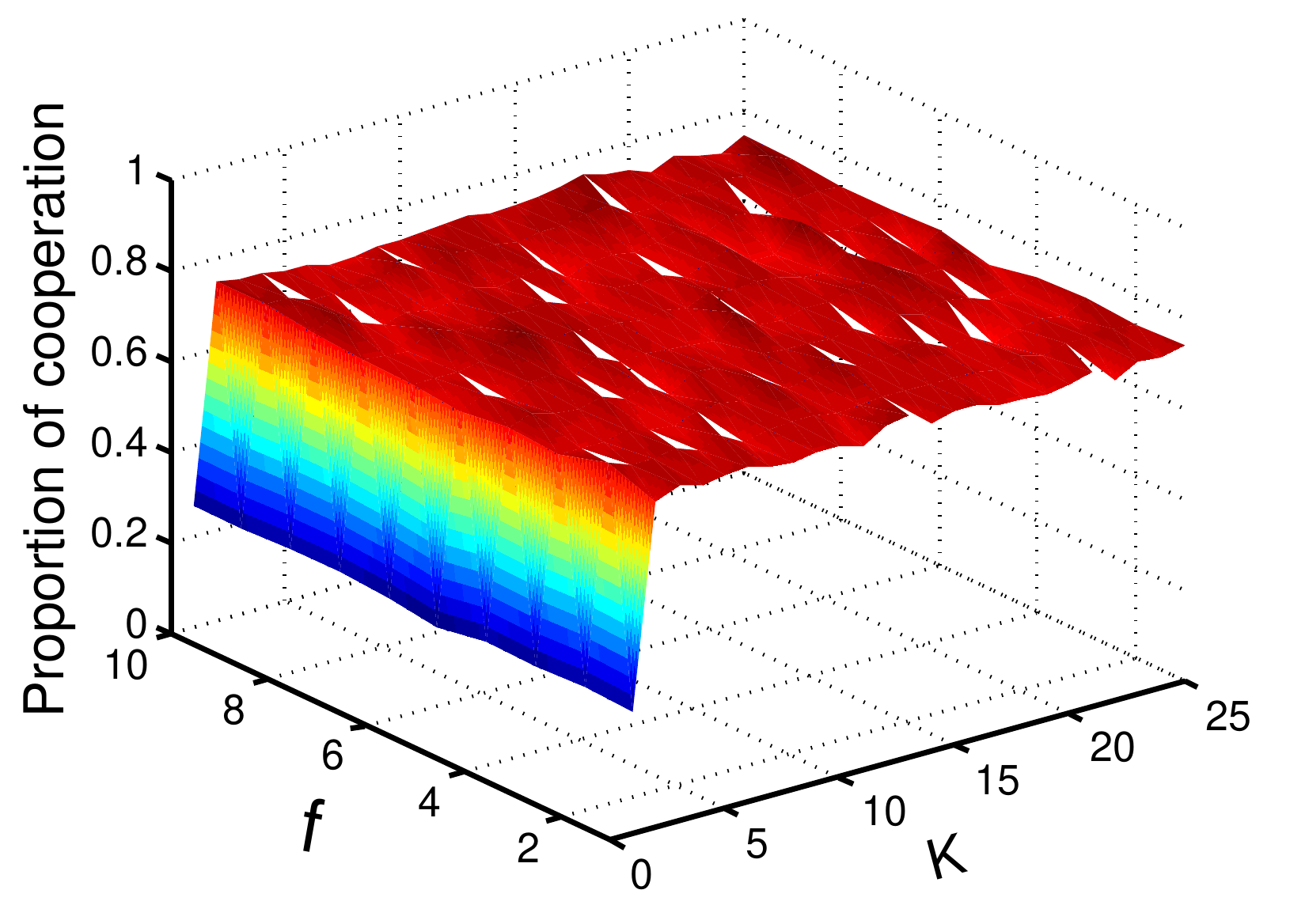}}
\subfloat[ZD deployment]{\includegraphics[width=1.9in,height=1.9in]{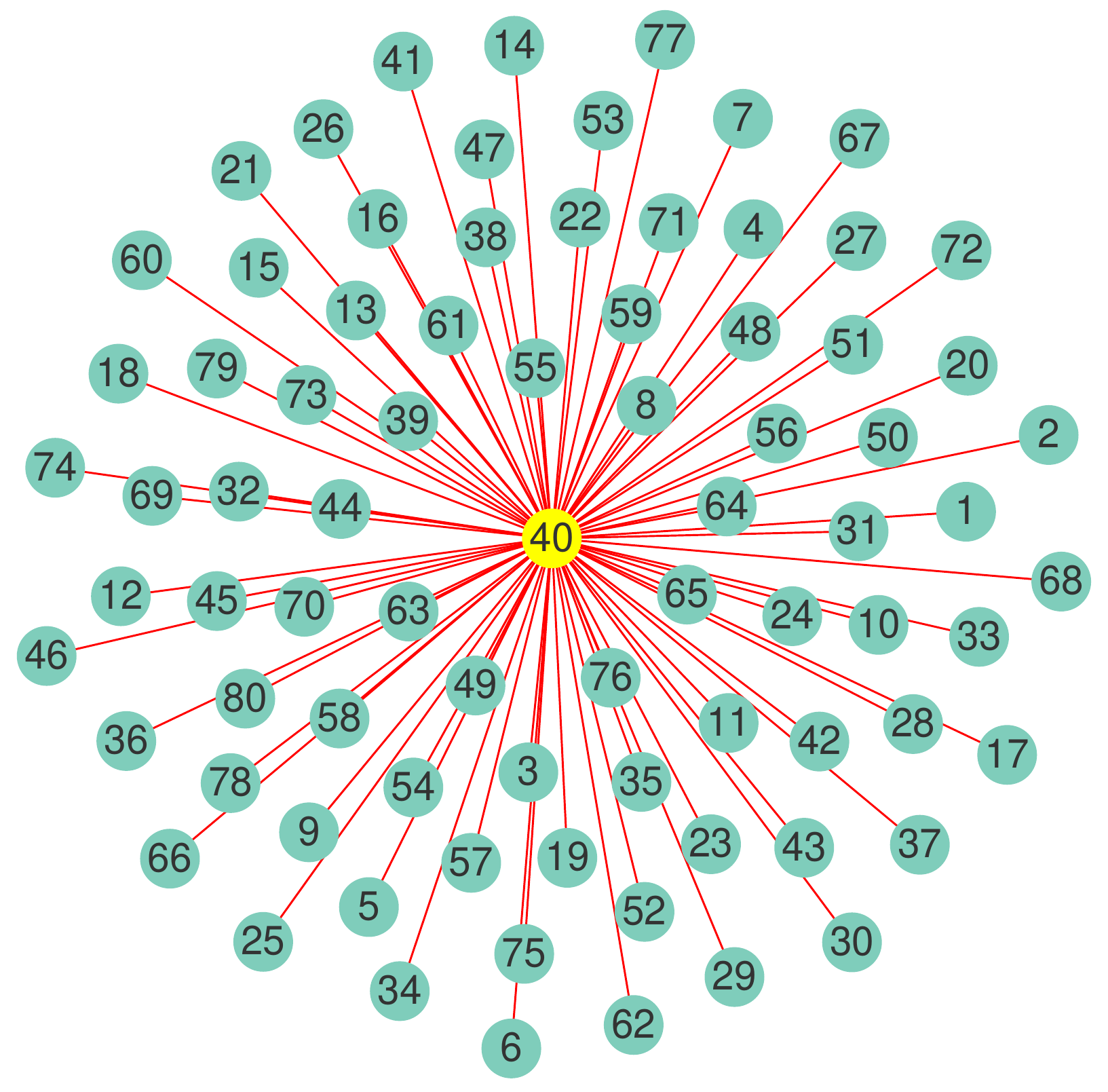}}
\hfill
\caption{Performance in the star  network.}
\label{fig:star}
\end{figure}

\begin{figure}[t]
\centering
\subfloat[Cooperator ratio]{\includegraphics[width=1.7in,height=1.6in]{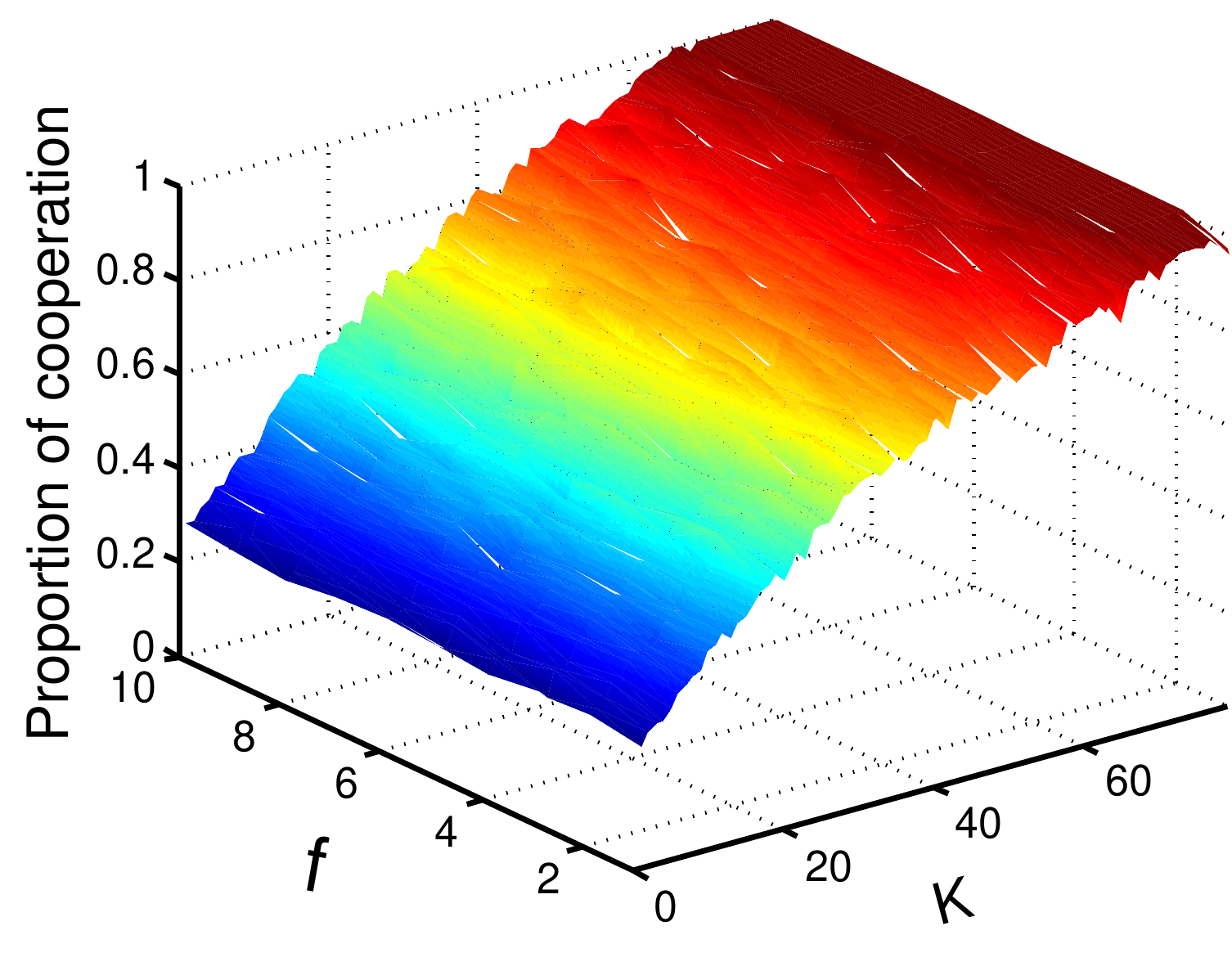}}
\subfloat[ZD deployment]{\includegraphics[width=1.9in,height=1.9in]{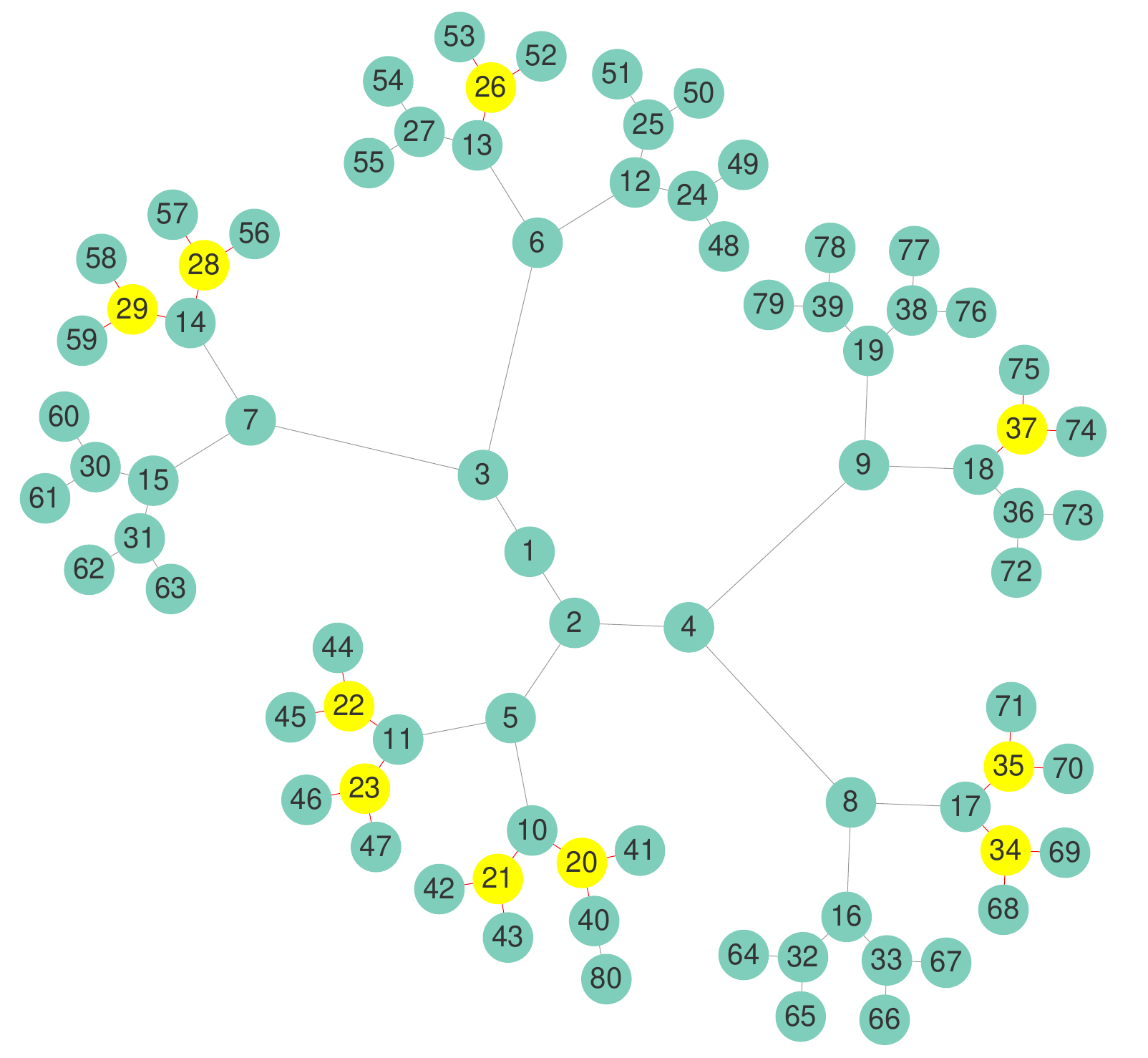}}
\hfill
\caption{Performance in the tree  network.}
\label{fig:tree}
\end{figure}

\begin{figure}[t]
\centering
\subfloat[Cooperator ratio]{\includegraphics[width=1.7in,height=1.6in]{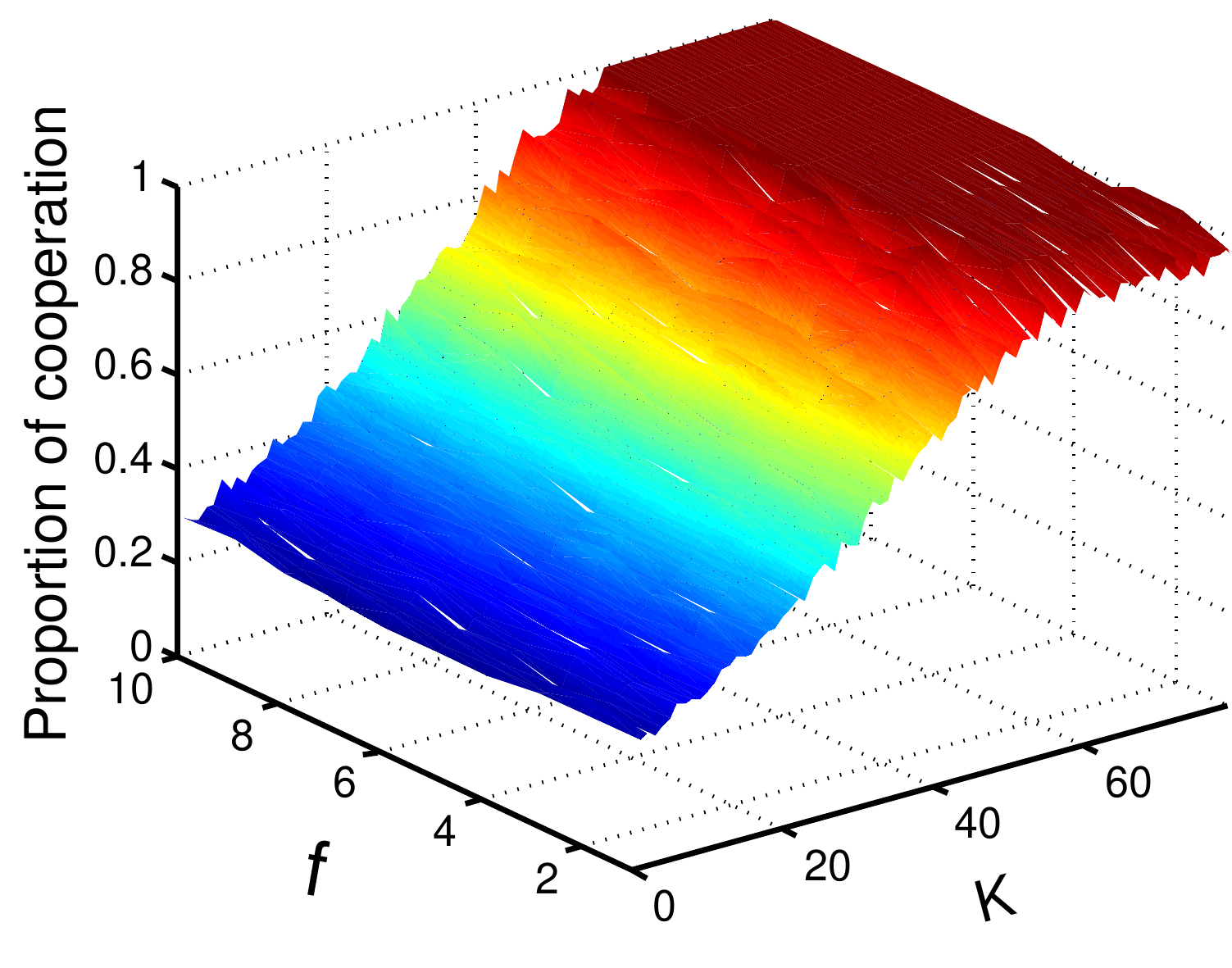}}
\subfloat[ZD deployment]{\includegraphics[width=1.9in,height=1.9in]{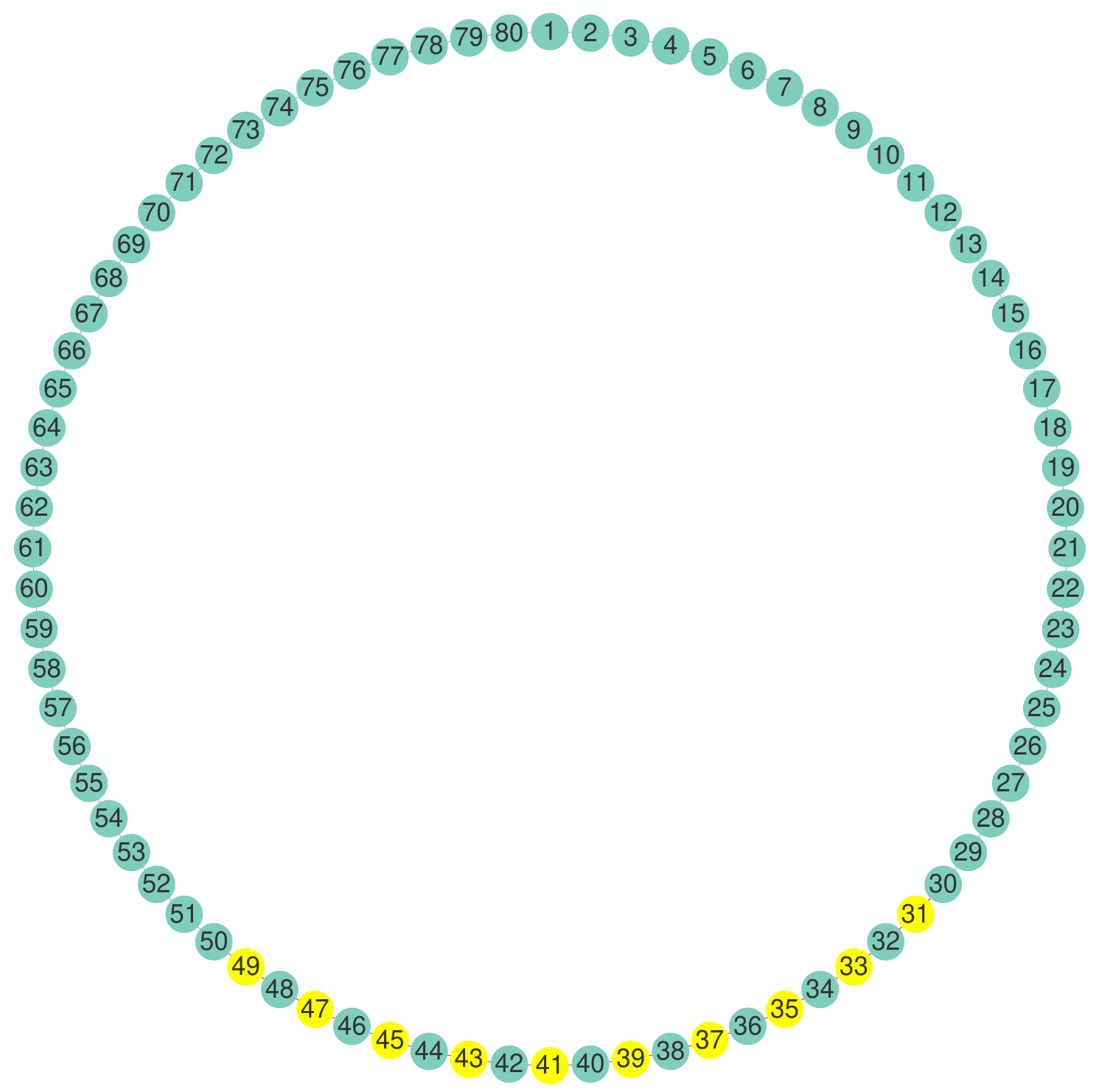}}
\hfill
\caption{Performance in the ring  network.}
\label{ring}
\end{figure}

Figs. \ref{fig:mesh}, \ref{fig:real}, \ref{fig:star}, \ref{fig:tree}, and \ref{ring} further demonstrate the performance of the proposed egoistic incentive mechanism for the mesh, real, star, tree, and ring networks, respectively, when $r=fn+3$, where $f$ is a coefficient. The subfigures (a) indicate the ratio of cooperators varying with $f$ and the number of ZD players $K$ in different networks while all subfigures (b) show the optimal deployments of ZD players in the corresponding networks when $f=2$ and $K=10$. The red lines in the subfigures (b) identify the social ties (edges) of the ZD players.

We can clearly observe  that {\it the emergence of cooperation} phenomenon happens in the mesh network from Fig. \ref{fig:mesh}(a). That is,  a large number of cooperators arise after more than 3 ZD players are deployed. When the number of ZD players is higher than 5, almost all regular players choose to cooperate. Fig. \ref{fig:mesh}(b) indicates the positions of 10 ZD players. The statistical data reveals that these ZD players are  popular nodes because their average degree  is 51.7, compared to 39, the average degree of all nodes in the mesh network.

Fig. \ref{fig:real}(a) demonstrates that the real network has similar performance as the mesh network. However, a further analysis on  Fig. \ref{fig:real}(b) indicates that the ZD players are not hot spots since their average degree is 16.8, which is lower than that of the whole network, i.e., 21.8. Contrarily,  the ZD strategy is adopted by the nodes with high betweenness whose average value is 67.8888, almost twice as much as the average betweenness of the whole system.  Here, the betweenness of a node is defined as the total number of times a node acts as a relay along the shortest path between any two other nodes. The reason behind this phenomenon may lie in that there exist sparse subnetworks in the real network that cannot be controlled by the popular nodes but can be regulated by nodes with a high  betweenness value because
betweenness can be  employed to evaluate a node's  contribution to the network connectivity, quantifying to what extent this node can impact on others.

Fig. \ref{fig:star}(a) indicates that there still exists  an  emergence of cooperation in the star network after deploying one ZD player. Just as we have expected,  this ZD player is located at the central place indicated in  Fig. \ref{fig:star}(b), which has the highest popularity as well as connectivity. As we have analyzed earlier, even though we put more ZD players in other positions, they would not function since there exist no regular players as their neighbors, leading to  low regulation power because no ZD alliance can be formed  in this case. Hence, we just label the position of the single ZD player in  Fig. \ref{fig:star}(b).

According to  Fig. \ref{fig:tree}(a), there is no emergence of cooperation phenomenon in the tree network. Moreover, the ratio of cooperators in the network rises steadily and relatively slowly due to the low average degree of the tree network.   From subfigure (b), one can see that all ZD players are deployed in the nodes with degree being 3, higher than the root node  and the leaf nodes whose degree values are respectively 2 and 1. Higher popularity makes ZD players possess more regular players as their neighbors, increasing their regulation range.

Fig. \ref{ring}(a) demonstrates that the performance of our mechanism in the ring network is similar to  that in the tree network: no  emergence of cooperation and low growth ratio of the number of cooperators. However, there exists a difference between them if we compare Fig. \ref{fig:tree}(a) with Fig. \ref{ring}(a). Specifically, when the number of ZD players is smaller than 30, the tree network outperforms the ring network; but  when there are more than 30 ZD players deployed, the ring network wins. The underlying reason can be found from Fig. \ref{fig:tree}(b) and Fig. \ref{ring}(b). The cause of the first phenomenon lies in that each ZD player in the tree network  has a degree of 3, which is higher than 2, the degree of a ZD player in  the ring network; the second phenomenon happens because  ZD alliances can be formed more easily in the ring network than in the tree network and when  the number of ZD players increases, the power of ZD alliance can stimulate more cooperation. For example, as indicated by Fig.\ref{fig:tree}(b) and Fig. \ref{ring}(b),  when 10 ZD players are  deployed to shape the behaviors of others, no ZD alliance exists in the tree network while several ZD alliances are formed in the ring network. For instance, in the ring network, node 35 can  ally with not only node 33 to regulate node 34,   but also node 37 to control node 36.

\section{Conclusion}
\label{sec:conclusion}
This paper proposes a mechanism to realize large-scale egoistic incentives via optimally deploying ZD players to reward cooperation and punish defection. We further derive a ZD alliance strategy  for  sequential multiple-player repeated games to speed up cooperation. Our approach has the  traits of statelessness and stability, making it scalable and suitable for large-scale systems.  The simulation results demonstrate that mesh and real networks can achieve the best performance where {\it the emergence of cooperation} phenomenon happens when only a few ZD players are deployed. Although such phenomenon also exists in a star network, its topological  trait makes only one ZD player function, pinning the cooperation probability to be $73\%$. The tree and ring networks perform the worst due to a low average degree. However, compared to the tree network, ZD alliances are more easily formed in the ring network, making it perform better as the number of  ZD players increases.


\section*{Acknowledgment}

The authors would like to thank the support from the National Natural Science Foundation of China under grants 61772080 and 61472044, and the National Science Foundation of the US under grants CNS-1704397 and IIS-1741279.

\bibliographystyle{IEEEtran}
\bibliography{IEEEabrv}

\begin{IEEEbiography}[{\includegraphics[width=1in,height=1.25in,clip,keepaspectratio]{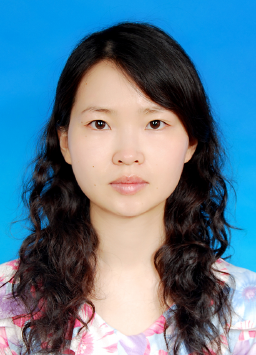}}]{Shengling Wang} is an associate professor in the College of Information
Science and Technology, Beijing Normal University. She received her Ph.D. in
2008 from Xi'an Jiaotong University. After that, she did her postdoctoral
research in the Department of Computer Science and Technology, Tsinghua
University. Then she worked as an assistant and associate professor from 2010 to 2013 in the Institute of Computing Technology of the Chinese
Academy of Sciences. Her research interests include mobile/wireless
networks, game theory, crowdsourcing.
\end{IEEEbiography}

\begin{IEEEbiography}[{\includegraphics[width=1in,height=1.25in,clip,keepaspectratio]{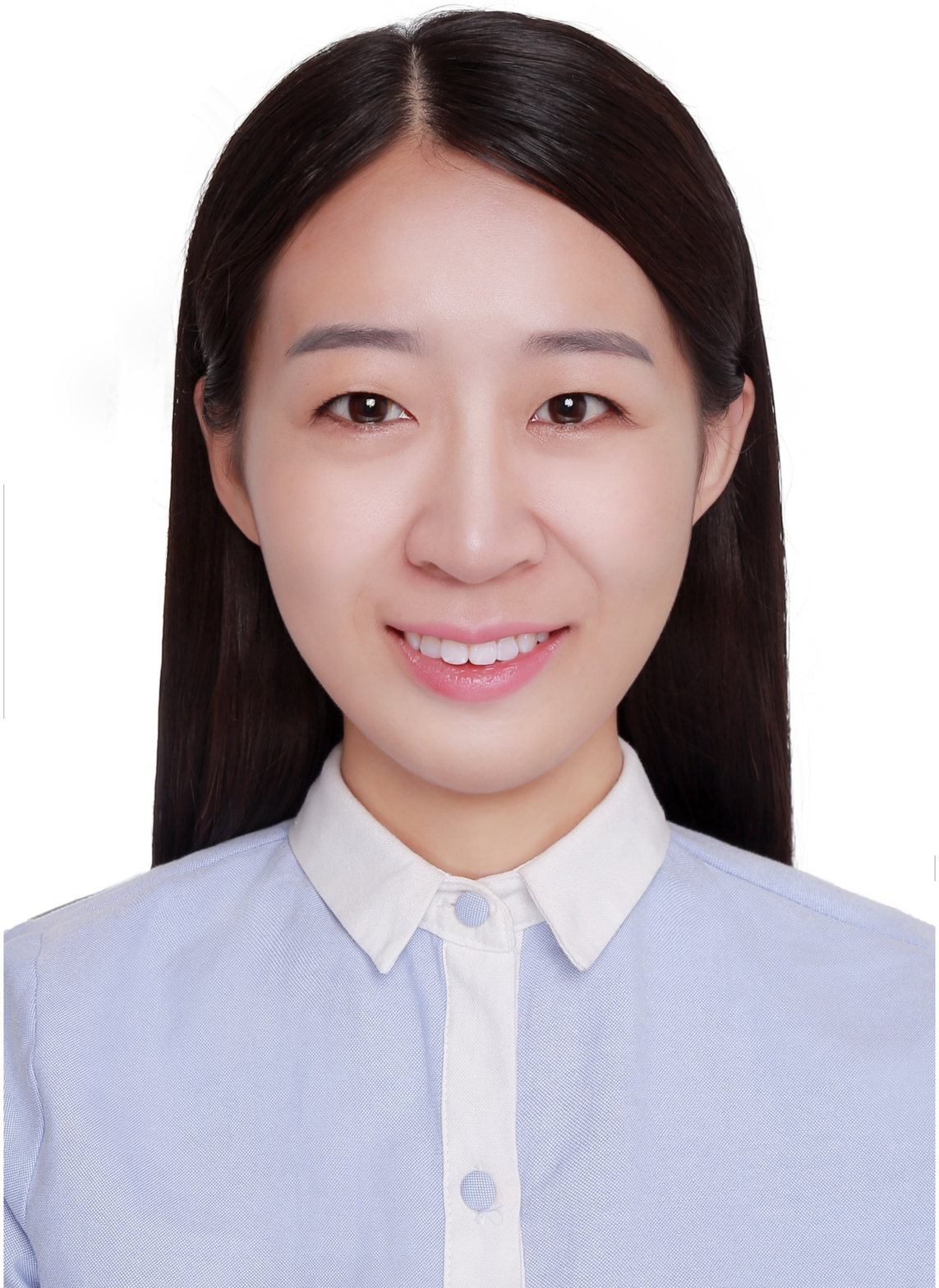}}]{Peizi Ma} received her B.S. degree in Computer Science and Technology
from Beijing Normal University in 2016. She is currently pursuing her M.S. degree in Computer Science at Beijing Normal University. Her research interests include wireless network, network science and game theory.
\end{IEEEbiography}

\begin{IEEEbiography}[{\includegraphics[width=1in,height=1.25in,clip,keepaspectratio]{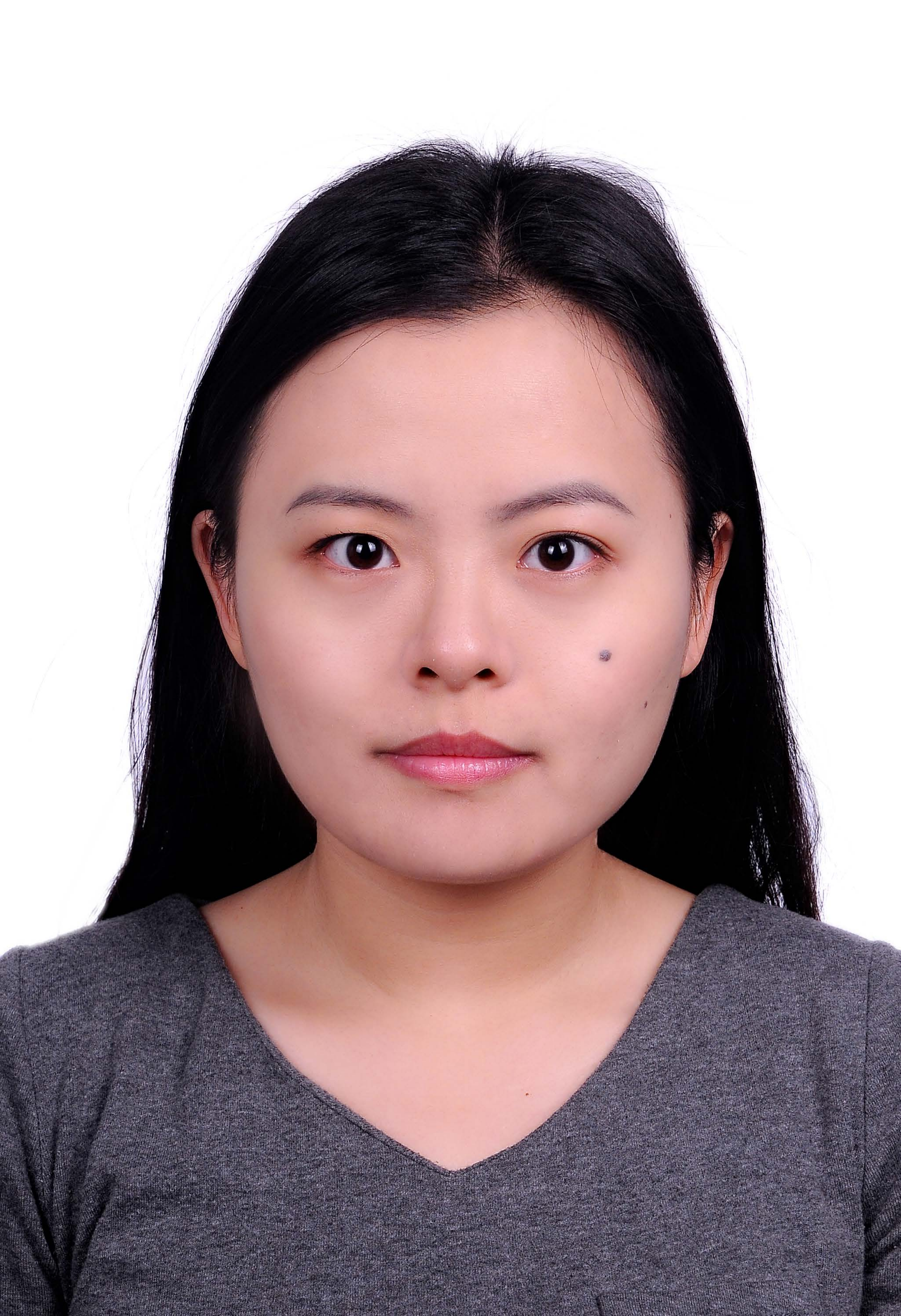}}]{Qin Hu} received her Ph.D. degree in Computer Science from the George Washington University in 2019. She is currently an Assistant Professor in the department of Computer and Information Science, Indiana University - Purdue University Indianapolis. Her research interests include wireless and mobile security, crowdsourcing/crowdsensing and blockchain.
\end{IEEEbiography}

\begin{IEEEbiography}[{\includegraphics[width=1in,height=1.25in,clip,keepaspectratio]{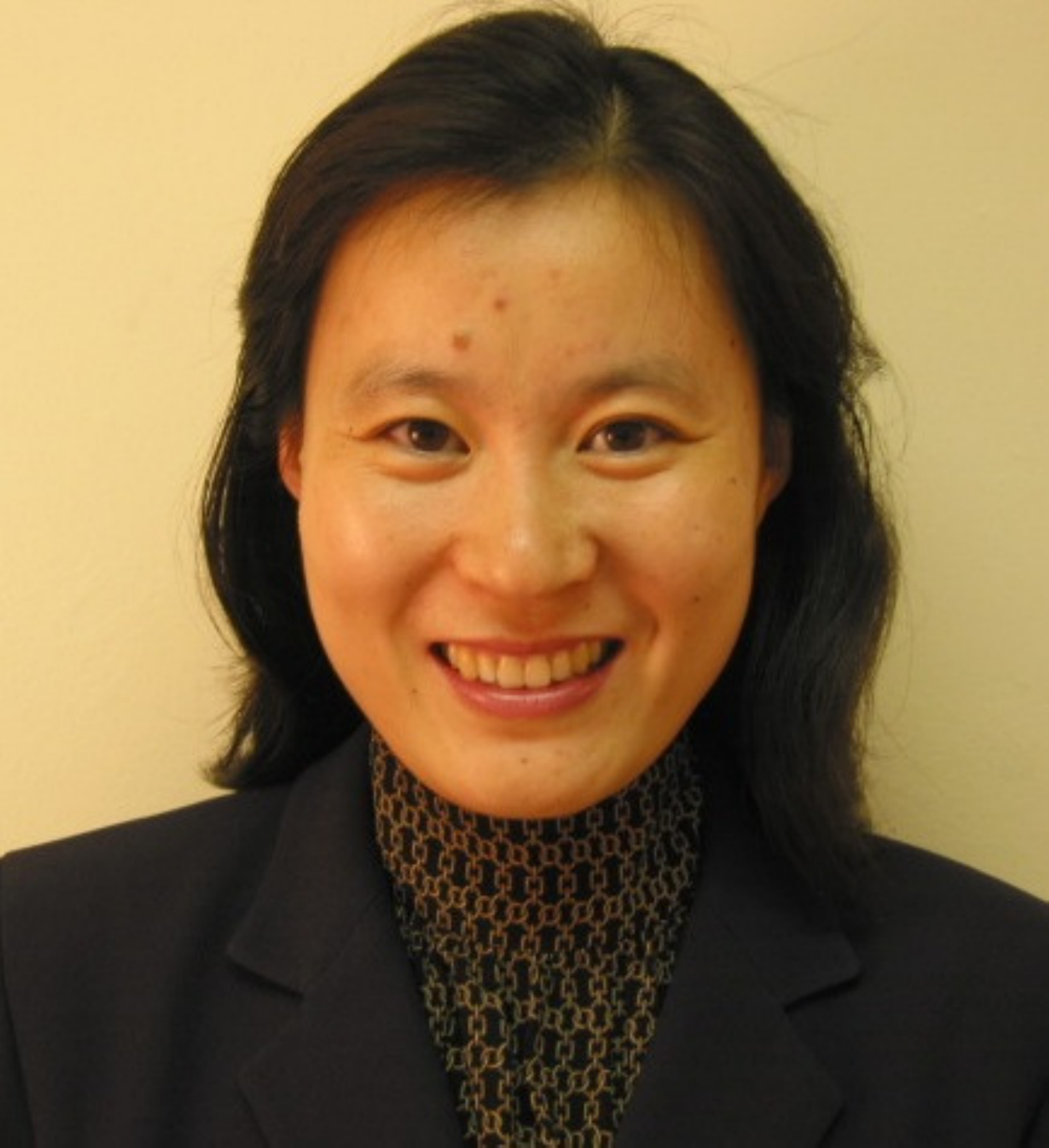}}]{Xiuzhen Cheng} [F] received her M.S. and Ph.D. degrees in computer
science from the University of Minnesota Twin Cities in 2000 and
2002, respectively. She is a professor in the Department of Computer
Science, George Washington University, Washington, DC. Her current
research interests focus on privacy-aware computing, wireless and
mobile security, dynamic spectrum access, mobile handset networking
systems (mobile health and safety), cognitive radio networks, and algorithm
design and analysis. She has served on the Editorial Boards of
several technical publications  and the Technical
Program Committees of various professional conferences/workshops. She
has also chaired several international conferences. She worked as a program director for the U.S. National Science
Foundation (NSF) from April to October 2006 (full time), and from
April 2008 to May 2010 (part time). 
\end{IEEEbiography}

\begin{IEEEbiography}[{\includegraphics[width=1in,height=1.25in,clip,keepaspectratio]{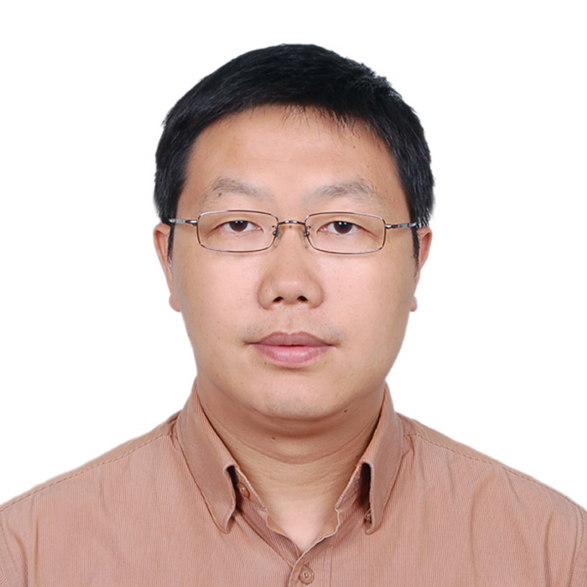}}]{Weifeng Lv} received his Ph.D. degree in computer science from Beihang University. His research interests include massive information system, urban cognitive computing, swarm intelligence, and smart cities. He is a professor of computer science, the dean of the School of Computer Science and Engineering, and the vice director of the State Key Laboratory of Software Development Environment, at Beihang University. He also serves as the Secretary-General of the China Software Industry Association and the director of National Engineering Research Center for Science and Technology Resources Sharing Service. He obtained multiple internationally renowned awards, including the second prize of the 2016 China National Science and Technology Invention Award and the first prize of the 2010 Beijing Science and Technology Award.
\end{IEEEbiography}

\end{document}